\documentclass[superscriptaddress,aps,amsmath,amssymb,showpacs,showkeys]{revtex4-2}
\usepackage{natbib}
\usepackage[dvips]{graphicx}
\usepackage{times}
\usepackage{braket}
\usepackage{xcolor}
\usepackage{orcidlink}
\usepackage{hyperref}
\usepackage{booktabs}
\usepackage{subfigure}
\usepackage{siunitx}
\hypersetup{
        colorlinks=true,
        urlcolor=magenta,
        linkcolor=red,
        citecolor=blue
}

\begin{document}

\title{Analysis of the Effect of Bars on Environmental Dependence of Disc 
Galaxies with MaNGA Survey Data}
	
\author{Pius Privatus\orcidlink{0000-0002-6981-717X}}
\email[Email: ]{privatuspius08@gmail.com}
\affiliation{Department of Physics, Dibrugarh University, Dibrugarh 786004, Assam, India}
\affiliation{Department of Natural Sciences, Mbeya University of Science and Technology, Iyunga 53119, Mbeya, Tanzania}
	
\author{Umananda Dev Goswami\orcidlink{0000-0003-0012-7549}}
\email[Email: ]{umananda@dibru.ac.in}
\affiliation{Department of Physics, Dibrugarh University, Dibrugarh 786004, Assam, India}
	
%\date{}
\begin{abstract}
Bars are fundamental structures in disc galaxies, although their role in 
galaxy evolution is still not fully known. This study investigates the effect 
of the presence of bars on the environmental dependence of disc galaxies'
properties using the volume-limited sample from Mapping Nearby Galaxies at 
APO (MaNGA) survey. The disc galaxies with and without bars samples were 
obtained using the Galaxy Zoo 2 project then assigned into isolated and 
non-isolated sub-samples. These sub-samples were used to compare the stellar 
mass, star formation rate, $g-r$ colour, concentration index and gas phase 
metallicity, and their relationships between isolated and non-isolated 
environments. Then these are used to investigate if there is an existence of 
any difference between galaxies with and without bars. A one-to-one 
correspondence between isolated and non-isolated galaxy properties 
was observed, and a strong dependence on the environment for properties of 
unbarred galaxies was observed when compared to barred. The stellar mass 
against star formation rate, $g-r$ colour against concentration index and 
stellar mass against gas phase metallicity of unbarred galaxies strongly 
depend on environment while for barred these relations weakly depend on 
environment. The study concludes that bars in disc galaxies decrease the 
dependence of analysed properties and their relations on the environment.
\end{abstract}
	
	
\maketitle    

\section{Introduction}
Bars are commonly seen in disc galaxies in today's Universe and are believed 
to play a significant role in shaping the development of galaxies by 
facilitating the transfer of angular momentum, within both the visible and 
dark matter components of the galaxy as suggested by simulations 
\cite{athanassoula2013bar,bournaud2002gas}. Hence they are very important 
structures in the disc galaxies, although their role in galaxy evolution 
is still not fully known \cite{cacho2014gaseous,sellwood2019global,
kruk2018galaxy,erwin2018dependence,deng2023dependence}. The mechanism for 
the growth of bars and the question as to why some galaxies have bars while 
others do not are still not understood in detail \cite{sellwood2019global,
bournaud2002gas,athanassoula2013bar}. Bars typically grow by capturing 
stars from the existing disc, they may also form stars from channelled 
gas; however, it is crucial to strike a balance between the growth of bars and 
the disc, as simulations indicate that excessive gas channelling could lead to 
bar destruction \cite{bournaud2002gas}. Ref.\ \cite{athanassoula2013bar} 
highlighted that a significant number of bars are believed to originate from 
disc instability and further noted that a disc with gas-poor content is more 
likely to develop a bar. 
	
Various studies have looked at how a bar impacts its host galaxy obtaining 
conflicting outcomes \cite{sellwood1993dynamics,george2021bar,
maeda2023statistical,kim2020effect,fraser2020sdss1, gavazzi2015halpha3,
jogee2006fueling,kormendy2004secular,knapen1995central,allard2006star,
ellison2011impact,carles2016mass,comeron2010ainur}. Bars are also effective 
in pushing gas towards the centre of galaxies to fuel star formation bursts 
where this process could potentially raise the metal content in the regions 
of both gas and stars \cite{sellwood1993dynamics,gavazzi2015halpha3}. Gas that 
accumulates in the areas of a galaxy due to the bar can be utilised for star 
formation as mentioned in Ref.\ \cite{jogee2006fueling} as well to fuel the 
central black hole or to enhance the central mass concentration 
\cite{kormendy2004secular}. While it remains uncertain whether bars 
directly cause gas consumption, indications point to a connection between 
bars and this phenomenon \cite{knapen1995central,comeron2010ainur,
allard2006star,ellison2011impact,fraser2020sdss1,sheth2005secular,
erwin2018dependence,coelho2011bars,kruk2018galaxy,ellison2011impact,
newnham2020h,carles2016mass}. There is evidence of an increased rate of 
forming stars at the centres of barred galaxies e.g.~see Ref.\ 
\cite{knapen1995central}, which often appear as nuclear rings 
\cite{comeron2010ainur}. The time for this enhanced rate of forming stars is 
still unknown although the estimate from the stellar population analysis 
indicates that these are short-lived and may consist of a sequence of 
continuous eruptions  \cite{allard2006star}. Ref.\ \cite{ellison2011impact}  
found that the metallicity in the central area is increased, however, fiber 
observations of low-mass and low-redshift galaxies showed no similar rise in 
star formation rate (SFR). According to the study by Ref.\ 
\cite{fraser2020sdss1}, barred galaxies have older and more metal-rich 
stellar populations when compared to unbarred. Since bars are observed to 
have both suppress and enhance the SFR in disc galaxies it is difficult to 
mention if bars are the driver of SFR or the result of star formation 
cessation in galaxies.
	
To determine whether there are differences between galaxies with and without 
bars, many observational studies have attempted to compare galaxies with and 
without bars, however, their findings are not conclusive 
\cite{sheth2005secular,erwin2018dependence,coelho2011bars, kruk2018galaxy,
ellison2011impact,newnham2020h}. Unbarred galaxies have a greater SFR than 
barred galaxies according to Ref.\ \cite{kruk2018galaxy}, which compares 
a volume-limited sample of barred galaxies matching the galaxies' stellar 
masses (M$\star$) with a sample of unbarred galaxies making a total of 
$\sim 3500$ galaxies with $z < 0.06$ from the Sloan Digital Sky Survey in 
order to investigate the relationship between disc features and bulge in the 
presence of a large galaxy bar. Authors discovered that although there is no 
significant difference in colours, the discs of galaxies without bars are 
noticeably bluer than those barred. Their analysis leads to the conclusion 
that this situation deserves further research, both theoretical and empirical. 
There is much evidence for gas flow along bars, including the increased 
central metallicity in spiral galaxies with bars when compared to unbarred 
spirals regardless of the global SFR results and higher central gas molecular 
content \cite{ellison2011impact,newnham2020h,sheth2005secular}. The fact that 
galaxies with bars are redder than those without bars and that the fraction 
of gas and SFR are lower at constant M$\star$ is possible evidence for the 
role of the bars in reducing the SFR as pointed out by Ref.\ 
\cite{kruk2018galaxy}, despite the fact that Ref.\ \cite{erwin2018dependence} 
observed that there is no evidence to confirm such a relationship.
	
Single-fibre analyses of the star population in barred galaxies' center area 
(bulge) reveal similarities to those of unbarred galaxies. Ref.\ 
\cite{cacho2014gaseous} states that there is no noticeable distinction 
in the metallicity (not gas or stars) of barred and unbarred galaxies. 
According to Ref.\ \cite{erwin2018dependence}, the host galaxy's M$\star$ 
affects the incidence of bars where for galaxies with mass 
M$\star\le 10^{9.7}$M$\odot$, the fraction of bars increases with M$\star$, 
while remain constant between 50\% and 60\% for more massive galaxies. The 
size of the bar also depends on the M$\star$ of the galaxy for which 
Ref.\ \cite{erwin2018dependence} found that there is a two-fold relationship 
between bar size and M$\star$: the size of the bar is nearly constant in 
galaxies at $\sim1.5$ kpc and M$\star\leq 10^{9.7}$M$\odot$, and the bar 
size at higher M$\star$ scales as $\propto$ M$\star^{0.56}$. 
Ref.\ \cite{lee2012dependence} investigating the dependence of bars occurrence 
on properties of galaxies and environments using a volume-limited sample 
generated from the seventh release of Sloan Digital Release Sky Survey, 
notice that the redder the $u-r$ colour, the increase of strong bar will be, 
and the highest value in the intermediate velocity dispersion. This trend 
indicates that medium mass systems often have strong bars. The low-mass, 
low-density blue galaxies are preferred for weak bars while the strong bars 
affect the concentration index of the galaxies when massive galaxies are 
considered. They again found that when other physical properties of the 
galaxy (such as $u-r$ colour) were fixed the bar fraction was not directly 
affected by the large background density. Furthermore Ref.\ 
\cite{lee2012dependence} discovered that for strong bars, the distance to the 
nearest neighbour galaxy reduces when the neighbour's virial radius drops to 
less than 10\%, regardless of the neighbour's morphology. These findings 
suggest that the mechanism underlying this phenomenon is gravitational 
rather than hydrodynamic. They further showed that strong bars collapse 
during the strong movement of the tides, then concluded that the fraction of 
weak bars was not related to environmental parameters. 
	
A study by Ref.\ \cite{lee2012bars} examining the relationship between bars 
and AGN activity in late-type galaxies, reveals that AGN host galaxies have 
a higher fraction of bars (42.6\%) compared to non-AGN galaxies (15.6\%). 
However, this trend is influenced by the known fact that AGN host galaxies 
are generally more massive and redder. When controlling for factors like 
colour and M$\star$, the differences between bar presence and AGN activity 
disappear, indicating that bars do not enhance AGN activity. These findings 
suggest that there is no conclusive evidence that bars in galaxies stimulate 
AGN activity. Presenting initial findings from Galaxy Zoo 2 
\cite{masters2011galaxy}, focusing on 13,665 disc galaxies to analyse the 
prevalence of bars in these galaxies based on various properties like colour 
and luminosity, Ref.\ 
\cite{masters2011galaxy} observed that approximately 29.4\% of the galaxies 
in the sample have bars, aligning with previous visual classifications but 
lower than automated methods. A noteworthy trend shows that the fraction of 
bars increases with a redder colour, less luminosity, and a dominated bulge 
indicating that more than half of the galaxy's bar has a mostly red bulge.  
The results indicate a colour dichotomy in disc galaxies where the red 
sequence is dominated by the bar and bulge while the blue clouds have minimal 
bulge or bar evidence. This supports theories of galaxy evolution, despite the 
fact that the results were discussed in the context of the inner evolution 
scenario in relation to the bar and bulge formation of disc galaxies.
	
The majority of previous research has mainly used single-fiber spectroscopy 
to explain the evolution of bars and how they are influenced by both internal 
and external factors in understanding galaxy evolution \cite{chacon2024bar,
thompson1981bar,sarkar2021galactic,skibba2012galaxy,byrd1990tidal,
gerin1990influence,eskridge2000frequency}. According to Ref.\ 
\cite{thompson1981bar}, the proportion of barred galaxies is higher in the 
Coma cluster's core than in the outer region. Ref.\ \cite{sarkar2021galactic} 
observed the absence of statistical correlation between the environment and the 
presence of bars in spiral galaxies. They further pointed out that bars are 
influenced by internal mechanisms larger than external, opposite to Ref.\ 
\cite{skibba2012galaxy}, presenting the influence of environmental dependence 
of bars and bulges in disc galaxies, which observed that the likelihood of 
having a bar increases as galaxies become massive and redder observing a 
significant bar-environment correlation. Ref.\ \cite{byrd1990tidal} shows how 
galaxies without bars can be shielded by the strong tidal isolated at the centre. 
Some studies using N-body simulations confirm that the bars are caused by 
external factors \cite{gerin1990influence,eskridge2000frequency}. According 
to Ref.\ \cite{eskridge2000frequency}, the fraction of bars in the Virgo and 
Fornax groups is slightly higher than the isolated, which implies that the 
fraction of bars may vary in different environments. This result emphasizes 
on the observational efforts to find relationships between environment and bar 
fractions \cite{eskridge2000frequency,aguerri2009population}.
	
Although the dynamical influence of bars has been studied in detail using 
individual cases \cite{thompson1981bar,byrd1990tidal,gerin1990influence,
eskridge2000frequency,aguerri2009population,debattista2019formation},
statistically large enough samples of barred galaxies from cosmological 
studies are essential for systematically understanding the bars. Moreover, it 
is clear that large samples of galaxies with bars from spatially resolved 
spectroscopy are necessary to carry out a detailed analysis of the stellar 
population gradient trend. To address these needs, the Mapping Nearby 
Galaxies survey at Apache Point Observatory (MaNGA) 
\cite{bundy2014overview, yan2016sdss,wake2017sdss} data and data 
from a citizen science effort (Galaxy Zoo 2) that attempts to distinguish 
light from structural components of galaxies within the MaNGA data cube 
have been implemented in this work. According to 
Ref.\ \cite{fraser2020sdss}, fractional bar lengths and main sequence of star 
formation are related, which uses a defined sample of 684 disc galaxies from 
the MaNGA survey to investigate the star formation and gas properties of a 
stretch of barred galaxies having various M$\star$ ranges and environments. 
The authors further concluded that star formation conditions within the core 
are controlled by shear, turbulence and wind flow. Therefore, the physical 
properties of the core are largely controlled by the host galaxy's available 
M$\star$ and it plays a role in the formation of stars in disc galaxies. 
	
Our goal in this study is to investigate if the presence of bars in disc 
galaxies influences the dependence of their physical properties (M$\star$, SFR, 
$g-r$ colour, r-band concentration index ($ci$) and gas phase metallicity) 
on the environment, using a large sample of local cosmic disc galaxies 
obtained from the integral field spectroscopy (IFS) data of the MaNGA 
survey \cite{bundy2014overview, yan2016sdss,wake2017sdss}. 
	
We present the survey and the methods of getting the samples in Section 
\ref{secII}. Section \ref{secIII} is used to present the results and in 
Section \ref{secIV} the results are discussed. We summarize the study in 
section \ref{secV} of this paper. We consider the standard cosmology with 
the Hubble constant $H_0=63$ km s$^{-1}$ Mpc$^{-1}$, matter density parameter 
$\Omega_{m}=0.3$, and dark energy density parameter $\Omega_{\Lambda}=0.7$. 
	
\section{Data} 
\label{secII}
\subsection{MaNGA survey}
The MaNGA survey \cite{bundy2014overview, yan2016sdss,wake2017sdss} 
is one of the fourth-generation Sloan Digital Sky 
Survey (SDSS) \cite{york2000sloan}, which has been used for the spatially 
resolved spectroscopic measurements studies containing $\sim 10^{4}$ galaxies 
employing $17$ fiber-Integrated Field Unit (IFU) operating in the range from 
$12$ to $32$ arcseconds ($19-127$ fiber per IFU) with a wavelength coverage 
of $3600-10300 \AA$, $R \sim 2000$ \cite{bundy2014overview,blanton2017sloan,
drory2015manga}. The target galaxies of MaNGA were chosen from a broad 
spectrum of various ranges in masses and colours within the redshift range of 
$0.01 < z < 0.15$. According to Refs.\ \cite{wake2017sdss,yan2016sdss}, 
the Primary + sample includes galaxies with a spatial coverage of up to 
$\sim1.5$ of galaxy effective radius (Re) for $\sim 66\%$ of the total sample. 
Re is representing the radius within which half of the total light of the 
galaxy is enclosed \cite{wake2017sdss,blanton2011improved}. The remaining 
(secondary) samples are typically seen at a higher redshift than the 
Primary + sample, out to $\sim 2.5$ Re. In this study, we use a volume-limited 
sample up to 
$z < 0.15$, with $500$ km/s line-of-sight velocity difference covering $1$ Mpc 
projected distance generated from the MaNGA survey, as detailed in 
Ref.\ \cite{abdurro2022seventeenth}, which characterizes the environment of 
large-scale structures  \cite{argudo2015catalogues}. The use of a 
volume-limited sample is very important for statistical purposes based on 
the fact that using a flux-limited sample introduces bias as the faint 
galaxies at large distances are obscured by the luminous galaxies  
\cite{teerikorpi2015eddington}.
	
\subsection{Barred and unbarred galaxy samples}
Morphological classifications for all MaNGA-restricted samples were obtained 
from the Galaxy Zoo 2 project as detailed in Refs.\ \cite{hart2016galaxy,
willett2013galaxy}. Galaxies' images were used by citizen scientists to 
identify if they are early, late, or merges, and additionally, more detailed 
features such as bars, bulges and edge shapes were measured. Respondents 
answered questions depending on their observation about the galaxy image, 
then were asked a follow-up question based on their answer using a decision 
tree mode at which Ref.\ \cite{willett2013galaxy} provides an unbiased 
opinion fraction for each of these questions. This has been improved in 
Ref.\ \cite{hart2016galaxy} with the limitations of considering good sample 
galaxies under each classification. It is very important to keep in mind that 
in this work, the limits recommended in Ref.\ \cite{willett2013galaxy} are 
used to find samples of barred and unbarred galaxies.  To ensure a robust 
selection of disc galaxies for the barred and unbarred 
samples, we applied the following set of well-defined morphological criteria:
	\begin{align}
		P_{features/disc}>0.430, \label{c1}\\[5pt]
		P_{notedgeon}>0.715, \label{c2}\\[5pt]
		N_{notedgeon}>20, \label{c3} \\[5pt]
		P_{bar}>0.8, \label{c4}\\[5pt]
		P_{bar}<0.2, \label{c5}
	\end{align}
where $P$ represents the probability of a particular galaxy type. 
First, galaxies were required to have a disc or feature probability 
greater than $0.430$, i.e., \( P_{\text{features/disc}} > 0.430 \) 
(Criterion~\ref{c1}), to preferentially select systems with prominent 
disc-like structures. To avoid complications due to edge-on orientations, 
which can obscure morphological features such as bars, only galaxies with a 
probability of not being edge-on exceeding $0.715$ were included, i.e., 
\( P_{\text{notedgeon}} > 0.715 \) (Criterion~\ref{c2}). Additionally, we 
required a minimum of $20$ votes from classifiers on the galaxy's orientation 
to ensure statistical reliability, leading to the condition 
\( N_{\text{notedgeon}} > 20 \) (Criterion~\ref{c3}). Barred galaxies were 
selected by imposing a high bar likelihood threshold of 
\( P_{\text{bar}} > 0.8 \) (Criterion~\ref{c4}), indicating strong visual 
evidence of a bar. Conversely, galaxies were classified as unbarred if 
they satisfied \( P_{\text{bar}} < 0.2 \) (Criterion~\ref{c5}), ensuring 
minimal probability of bar features. These criteria collectively enable the 
construction of clean and reliable samples for comparative morphological 
studies.

\begin{figure}[h!]
  \subfigure{
     \includegraphics[width=0.3\linewidth]{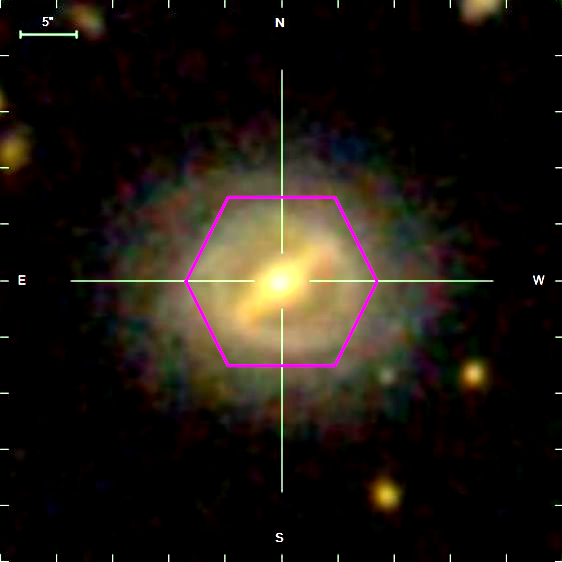}
  }
	\hspace{0.1cm}
	\subfigure{
	   \includegraphics[width=0.3\linewidth]{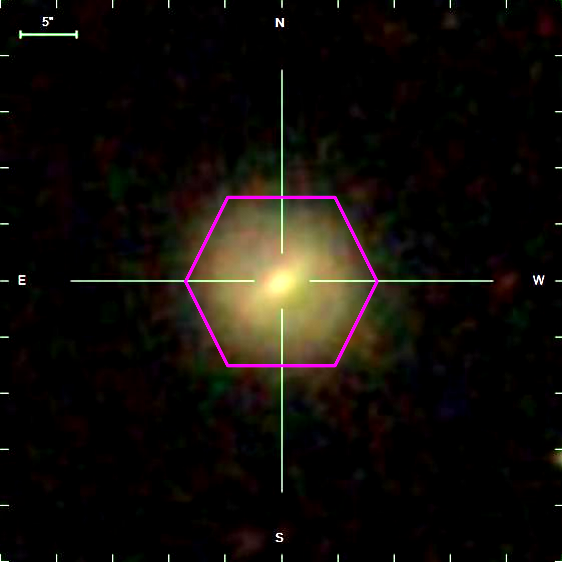}
	}
	\hspace{0.1cm}
	\subfigure{
	   \includegraphics[width=0.3\linewidth]{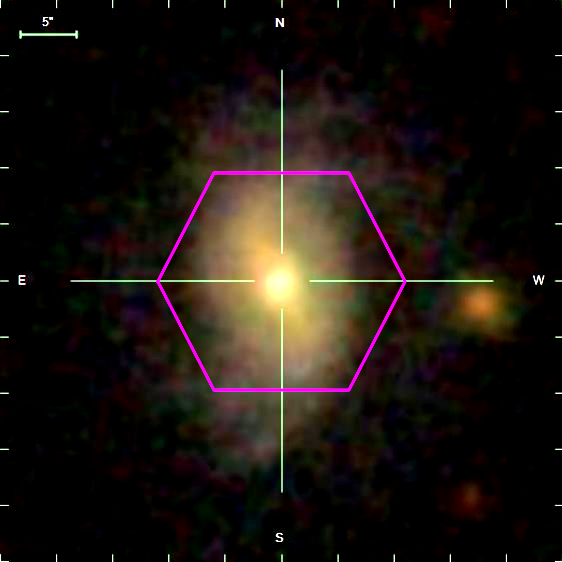}
	 }
	 \hspace{0.1cm}
	 \subfigure{
	    \includegraphics[width=0.3\linewidth]{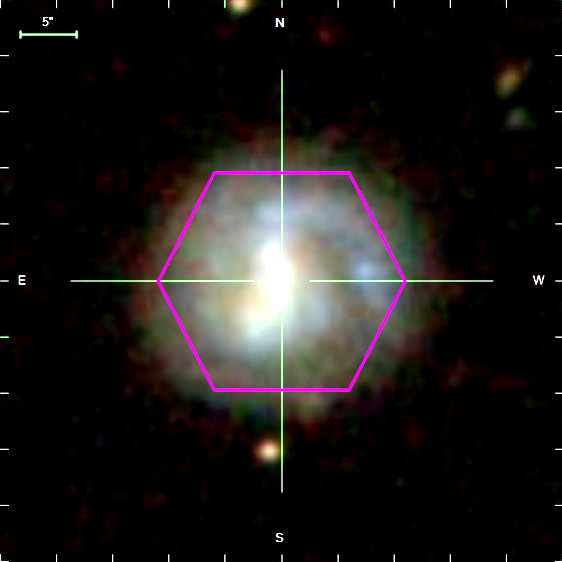}
	 }
	 \hspace{0.1cm}
	 \subfigure{
	     \includegraphics[width=0.3\linewidth]{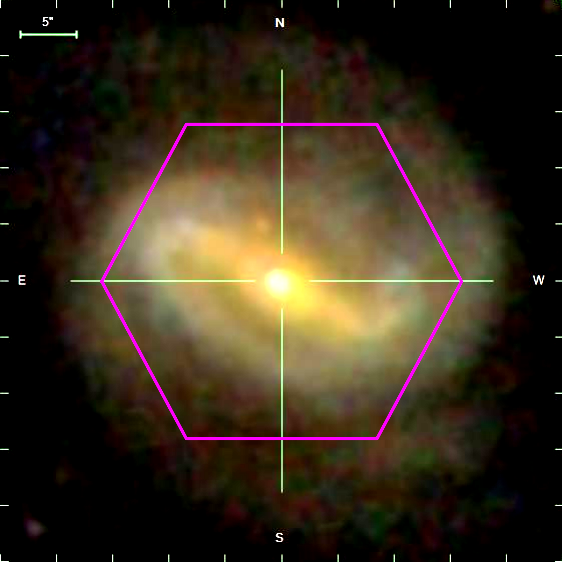}
	 }
	 \hspace{0.1cm}
	 \subfigure{
	     \includegraphics[width=0.3\linewidth]{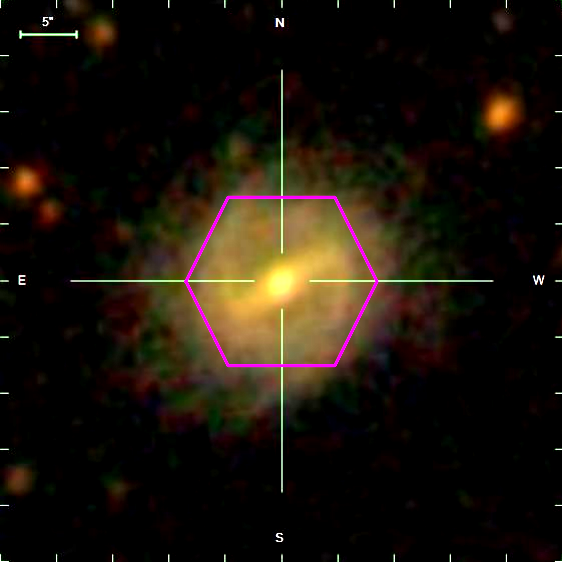}
	 }
\vspace{-0.2cm}
\caption{The images for $6$ out of $356$ barred galaxies denoted by their 
MaNGA plate and IFU, MaNGA$10226-3704$ (top left), MaNGA$9873-3701$ (top 
middle), MaNGA$10222-6103$ (top right), MaNGA$9876-6101$ (bottom left), 
MaNGA$9881-12705$ (bottom middle) and MaNGA$9513-3703$ (bottom right). The 
pink hexagon covers the spatial extent of the MaNGA Integral Field Unit.
}
\label{B}
\end{figure}

\begin{figure}[h!]
    \subfigure{
	 \includegraphics[width=0.3\linewidth]{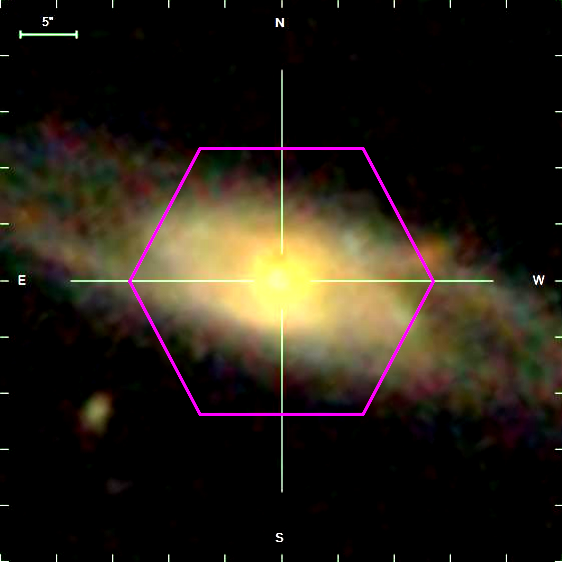}
    }
    \hspace{0.1cm}
    \subfigure{
	 \includegraphics[width=0.3\linewidth]{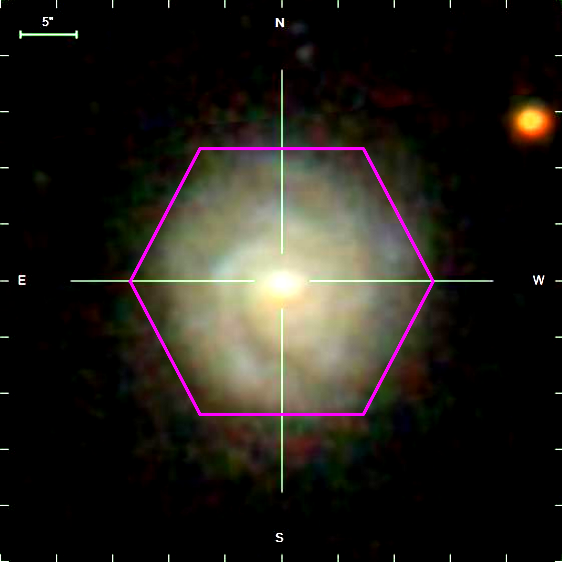}
    }
    \hspace{0.1cm}
    \subfigure{
	 \includegraphics[width=0.3\linewidth]{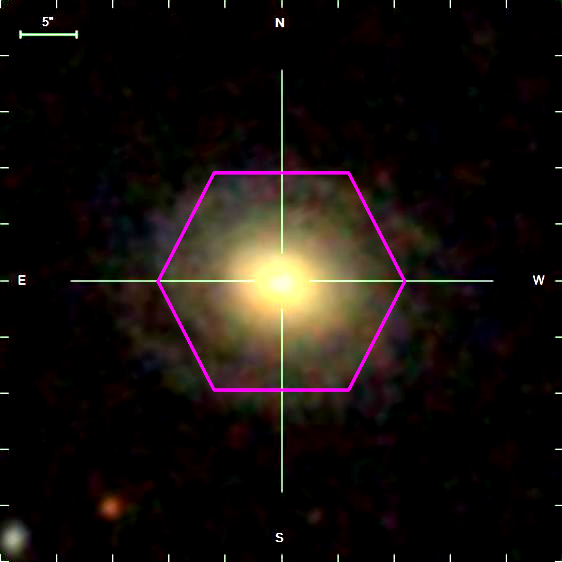}
    }
    \hspace{0.1cm}
    \subfigure{
         \includegraphics[width=0.3\linewidth]{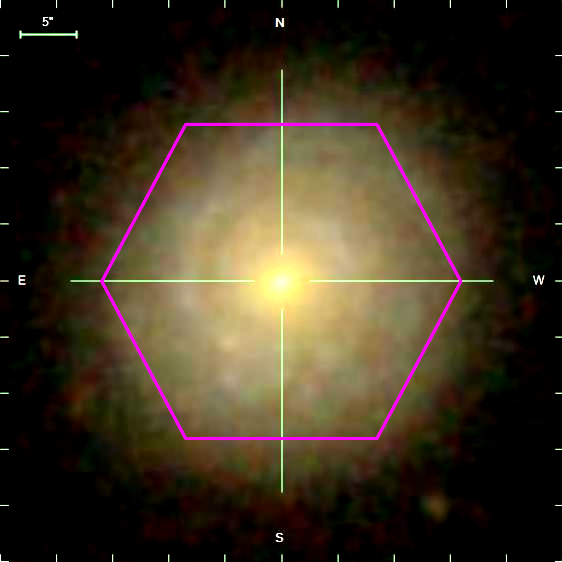}
    }
    \hspace{0.1cm}
    \subfigure{
	 \includegraphics[width=0.3\linewidth]{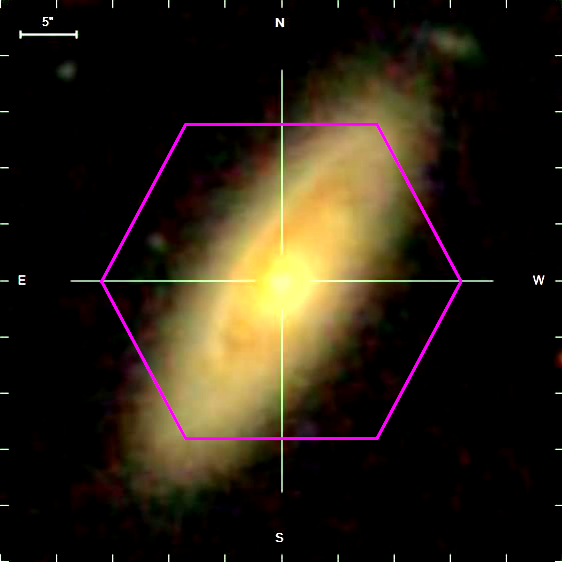}
    }
    \hspace{0.1cm}
    \subfigure{
	 \includegraphics[width=0.3\linewidth]{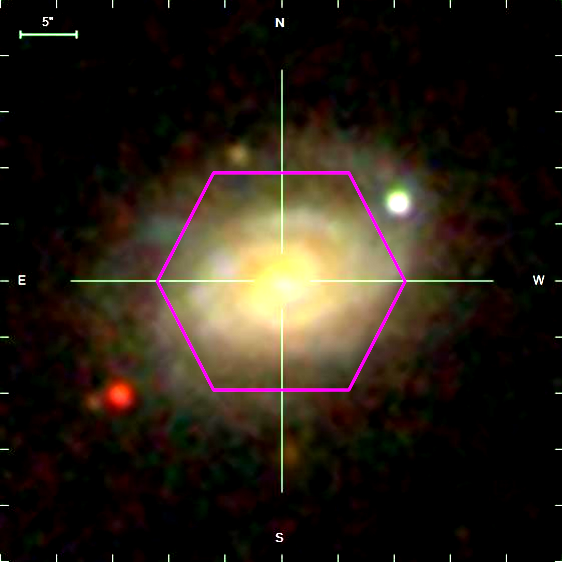}
    }
\vspace{-0.2cm}
\caption{The images for $6$ out of $1180$ unbarred galaxies denoted by their 
MaNGA plate and  IFU, MaNGA$10213-9101$ (top left), MaNGA$9879-9101$ (top 
middle), MaNGA$9890-6102$ (top right), MaNGA$8951-12705$ (bottom left), 
MaNGA$11939-12705$ (bottom middle) and MaNGA$9894-6104$ (bottom right). The 
pink hexagon covers the spatial extent of the MaNGA Integral Field Unit.
}
\label{NB}
\end{figure}

Hence for barred galaxies we employed criteria \eqref{c1}, \eqref{c2}, 
\eqref{c3}, and \eqref{c4} to select a total number of 356 barred galaxies.
For unbarred galaxies we employed criteria \eqref{c1}, \eqref{c2}, \eqref{c3}, 
and \eqref{c5} to make a total of 1180 unbarred galaxies. The samples of MaNGA 
images for barred galaxies are shown in Fig.~\ref{B} and unbarred galaxies 
in Fig.~\ref{NB}.

\subsection{Galaxy environment}
The Galaxy Environment for MaNGA Value Added Catalogue (GEMA-VAC), 
was used to quantify the barred and unbarred galaxies' environment, the 
details of GEMA-VAC will be provided in Argudo-Fernádez et al. (in prep.). 
This volume-limited 
value-added catalogue contains environmental quantification for several MaNGA 
galaxies, based on the methods described in Refs.\ \cite{argudo2015catalogues,
etherington2015measuring,wang2016elucid}. We use the information provided in 
the GEMA-VAC at which the galaxies are assigned in groups using a halo-based 
group finder by Ref.\ \cite{yang2007galaxy}. The galaxies which are alone in 
the group (Group size (GS) $=1$) are named as the isolated and the galaxies 
with at least one neighbour in the group (Group size (GS) $\geq2$) are named 
non-isolated galaxies. Remembering that galaxies selected in different 
environments may have significantly different stellar masses, that 
any observed changes in the other properties may be of stellar mass related. 
To mitigate this issue we have mass-matched all the samples in the stellar 
mass range of 8.5 $< \log_{10}$ M$\star$ $<$11.5. With these conditions 
applied to barred and unbarred pre-defined criteria a total number of $158$ 
(4$4.38\%$) and 198 ($55.62\%$) isolated and non-isolated barred galaxies were 
obtained. Furthermore, a total number of $572$ ($48.47\%$) and $608$ 
($51.53\%$) isolated and non-isolated unbarred galaxies were obtained. These 
samples are used in the next sections to compare the M$\star$, SFR, $g-r$ 
colour, $ci$ and gas phase metallicity between isolated and non-isolated 
environments. 
	
The parameter distributions from our sample were compared using the 
Anderson–Darling two-sample statistical test 
\cite{anderson1952asymptotic,pettitt1976two,scholz1987k} by taking the null 
hypothesis that all isolated and non-isolated samples originate from the same 
population \cite{scholz1987k}. The Anderson–Darling statistical test is 
preferred due to its more sensitive statistical results 
\cite{babu2006astrostatistics}. In principle, if isolated and non-isolated 
samples do not reject the null hypothesis, this indicates that our isolated 
and non-isolated 
samples are galaxies with the same population, so there is no dependency on 
the environment. The approach is outlined in Ref.\ \cite{sanchez2022sdss}, 
which processes IFS data cubes to extract spectroscopic properties that were 
used to obtain the galaxy physical properties used in this work. In addition 
to the parameters obtained from pyPipe3D \cite{sanchez2022sdss}, a set of 
photometric structure features can also be obtained directly from the MaNGA 
data cube such as broadband photometry in the B, V, R  and u, g, r, i filters. 
Filter parameters were used from Ref.\ \cite{fukugita1995galaxy}, which 
employed the Vega photometric system that redshifted to the rest frame of 
each object. 
	
\section{Results}\label{secIII}
\subsection{Galaxy properties}
\begin{figure}[h!]
    \subfigure{
       \includegraphics[width=0.46\linewidth]{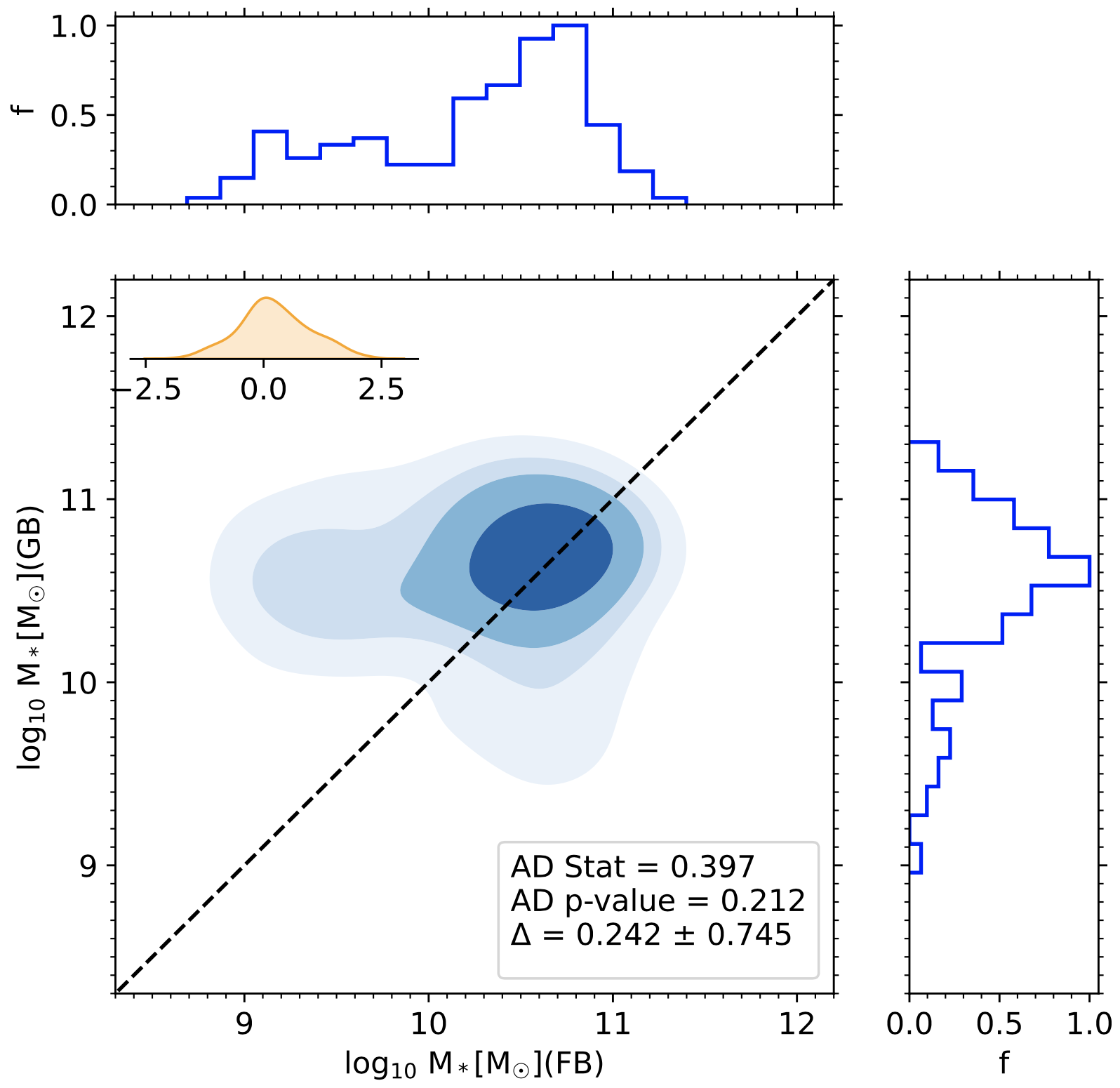}
    }
    \hspace{0.3cm}
    \subfigure{
	\includegraphics[width=0.46\linewidth]{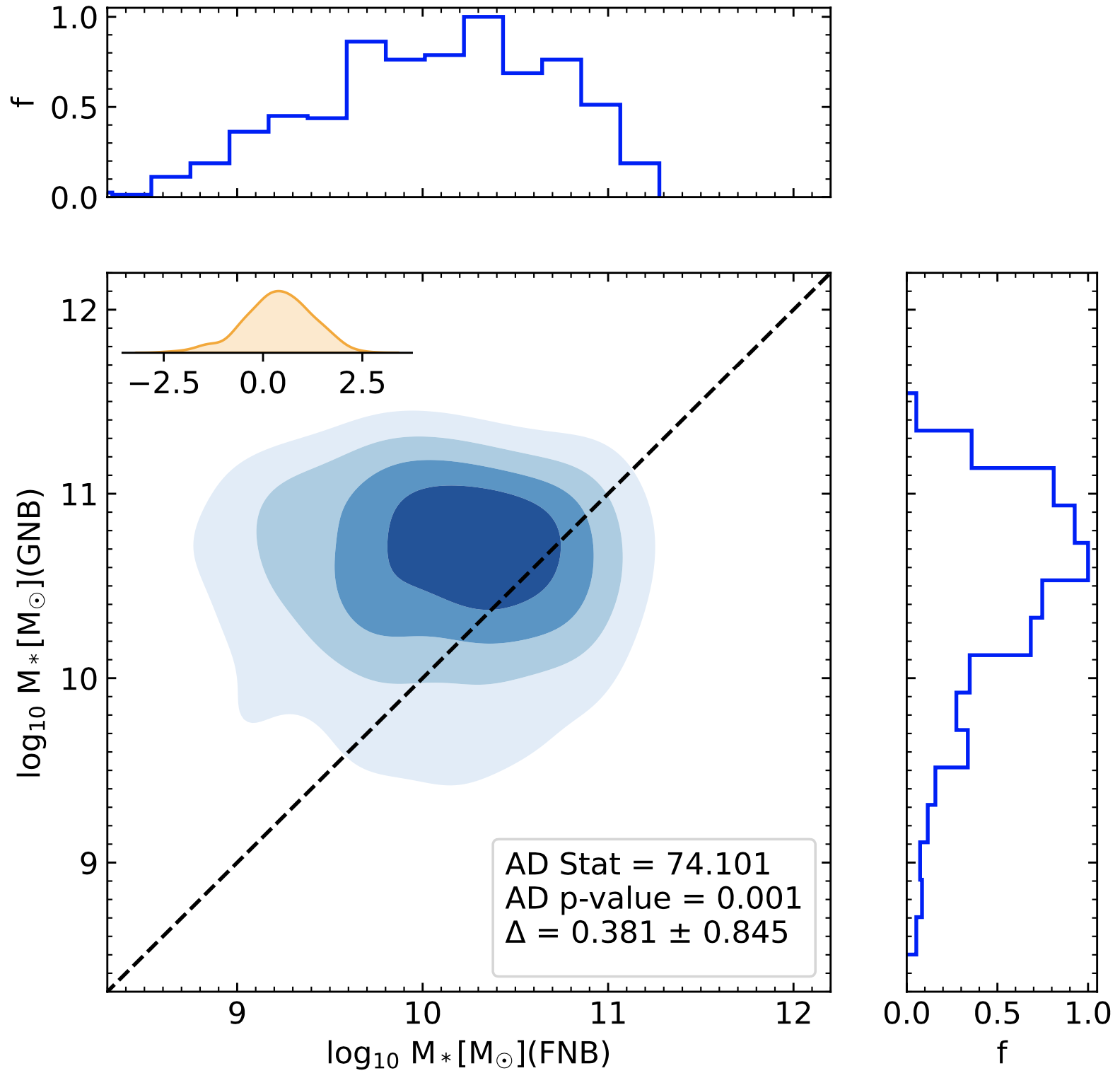}
    }
    \vspace{-0.2cm}
\caption{Comparison between stellar mass distributions of barred (left panel) 
and unbarred (right panel) isolated (x-axis) and non-isolated (y-axis) 
galaxies. In plots, GB stands for barred non-isolated galaxies, GNB for 
unbarred non-isolated galaxies, FB for barred isolated galaxies, FNB for 
unbarred isolated galaxies and AD for Anderson-Darling. The orange upper left 
inset plot represents the distribution of the differences between the isolated 
and non-isolated samples, where the x-axis is in dex while the y-axis shows 
the probability density, which is unitless. These terminologies and 
representations are used for the rest of the similar figures.   
}
\label{M}
\end{figure}

\begin{figure}[h!]
     \subfigure{
	\includegraphics[width=0.46\linewidth]{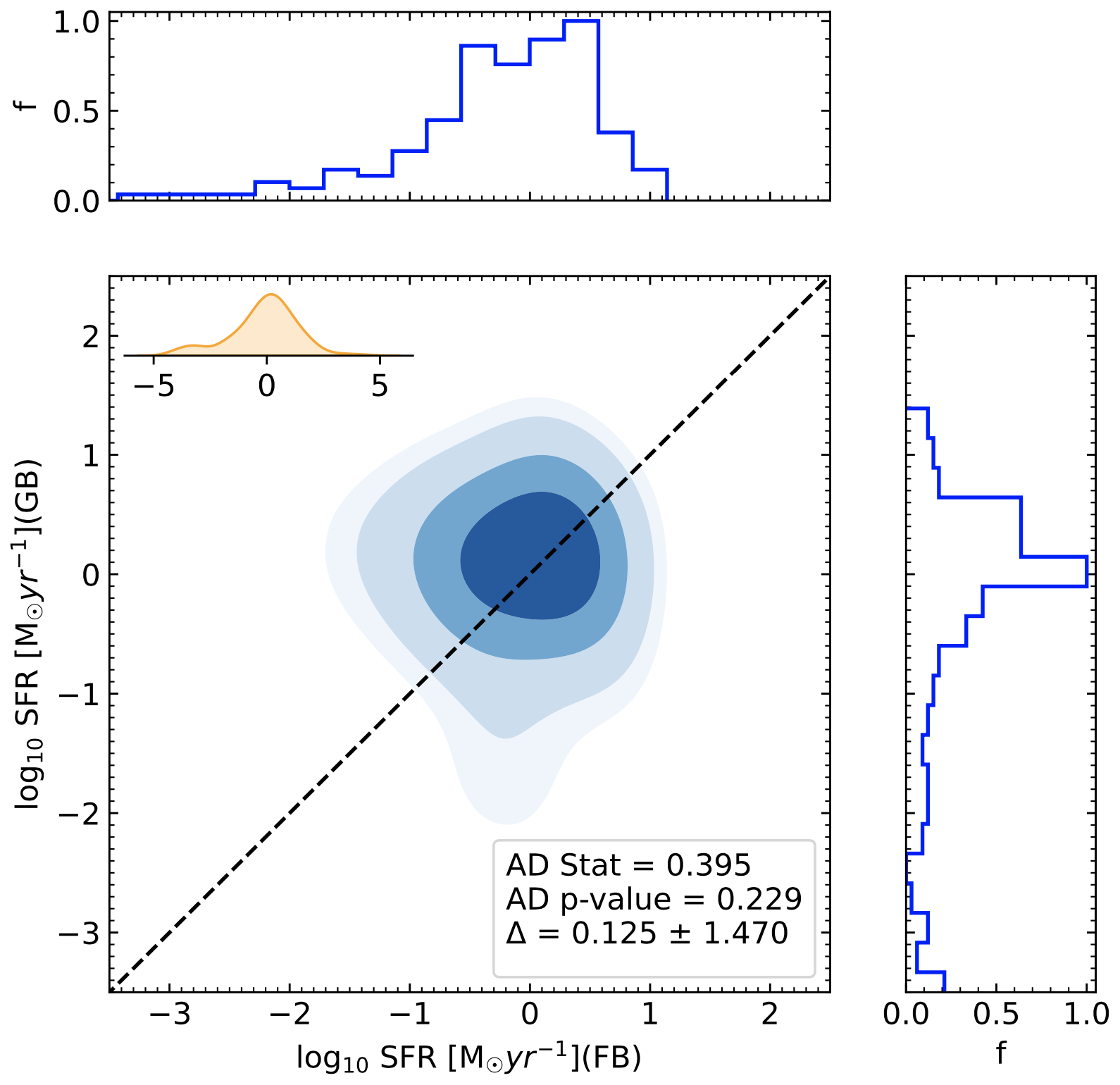}
     }
     \hspace{0.3cm}
     \subfigure{
	\includegraphics[width=0.46\linewidth]{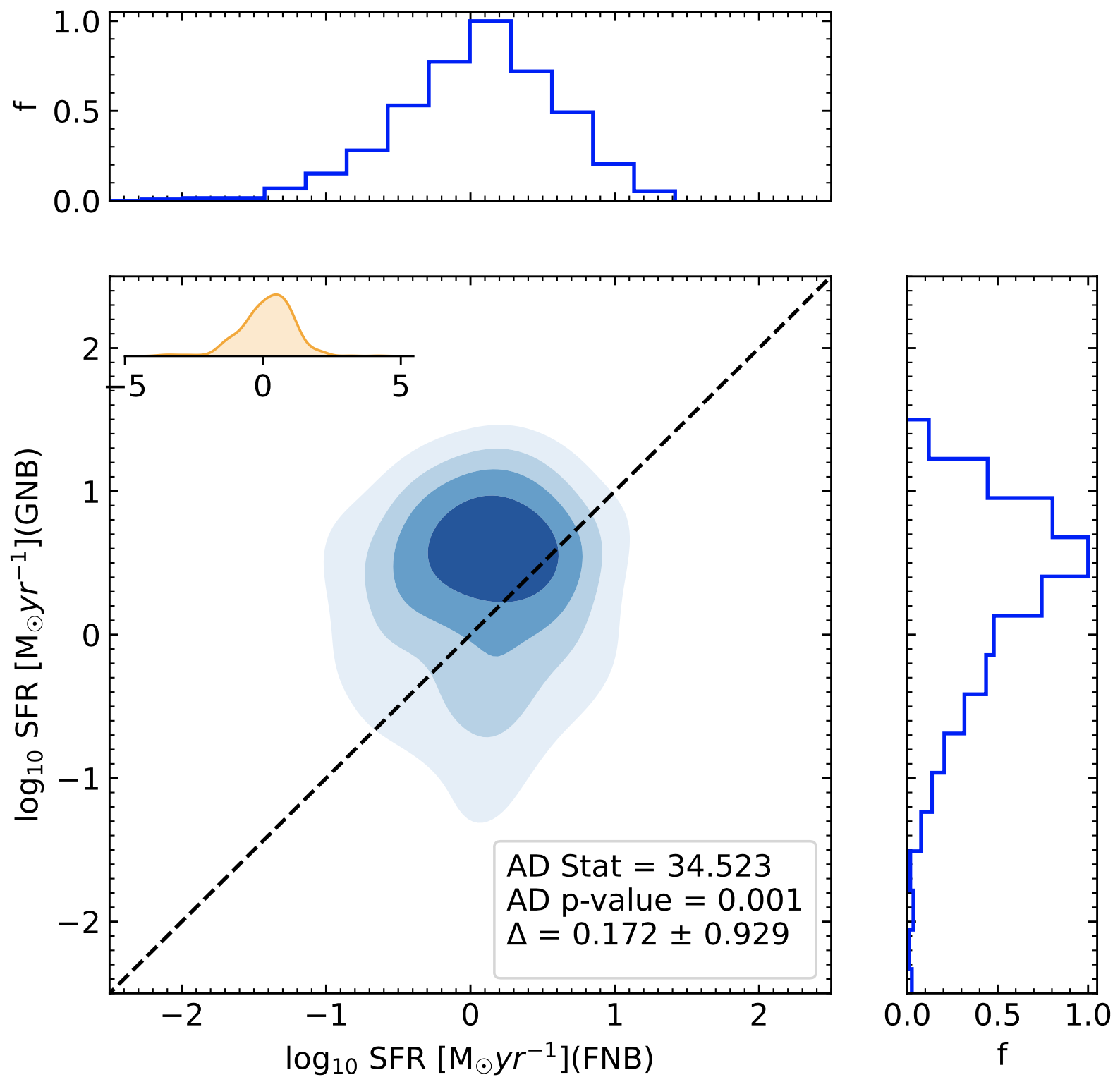}
     }
     \vspace{-0.2cm}
\caption{Comparison between SFR of barred (left panel) and unbarred (right 
panel) isolated (x-axis) and non-isolated (y-axis) galaxies.}
\label{SF}
\end{figure}

\begin{figure}[h!]
	\subfigure{
	        \includegraphics[width=0.46\linewidth]{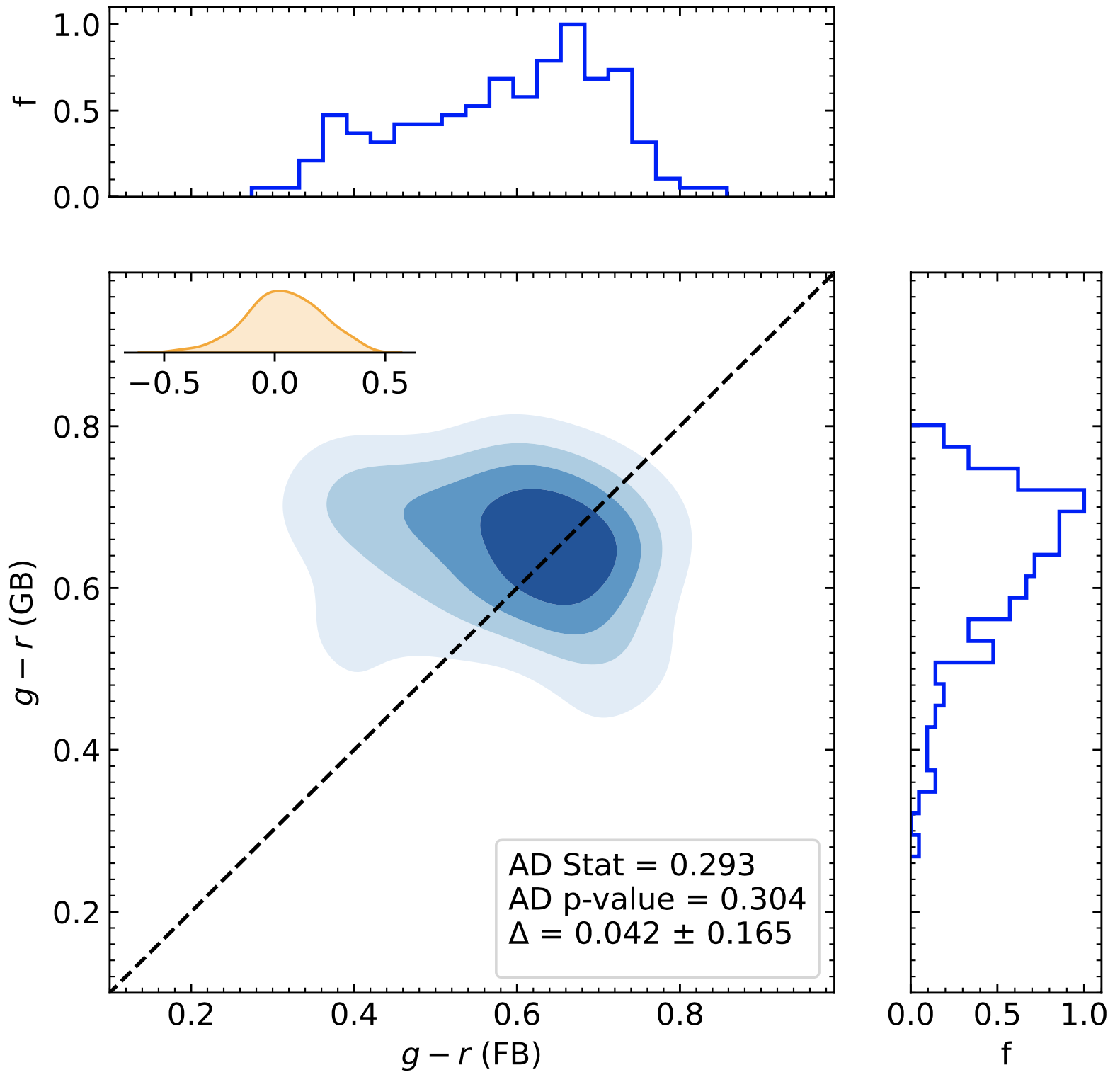}
	}
	\hspace{0.3cm}
	\subfigure{
		\includegraphics[width=0.46\linewidth]{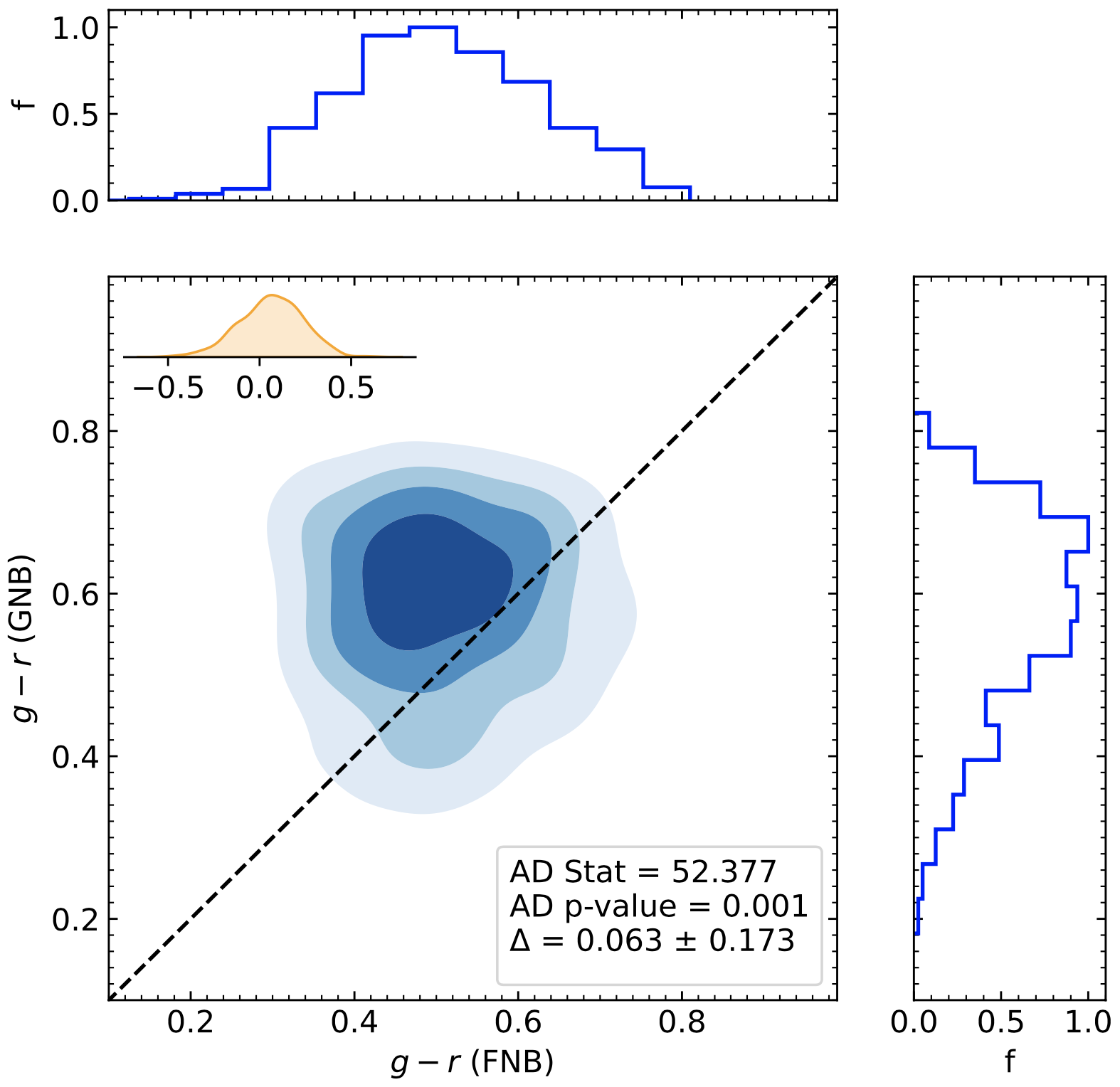}
	}
	\vspace{-0.2cm}
\caption{Comparison between $g-r$ colour distributions of barred (left panel) 
and unbarred (right panel) isolated (x-axis) and non-isolated (y-axis) galaxies.}
\label{GR}
\end{figure}

\begin{figure}[h!]
	\subfigure{
		\includegraphics[width=0.46\linewidth]{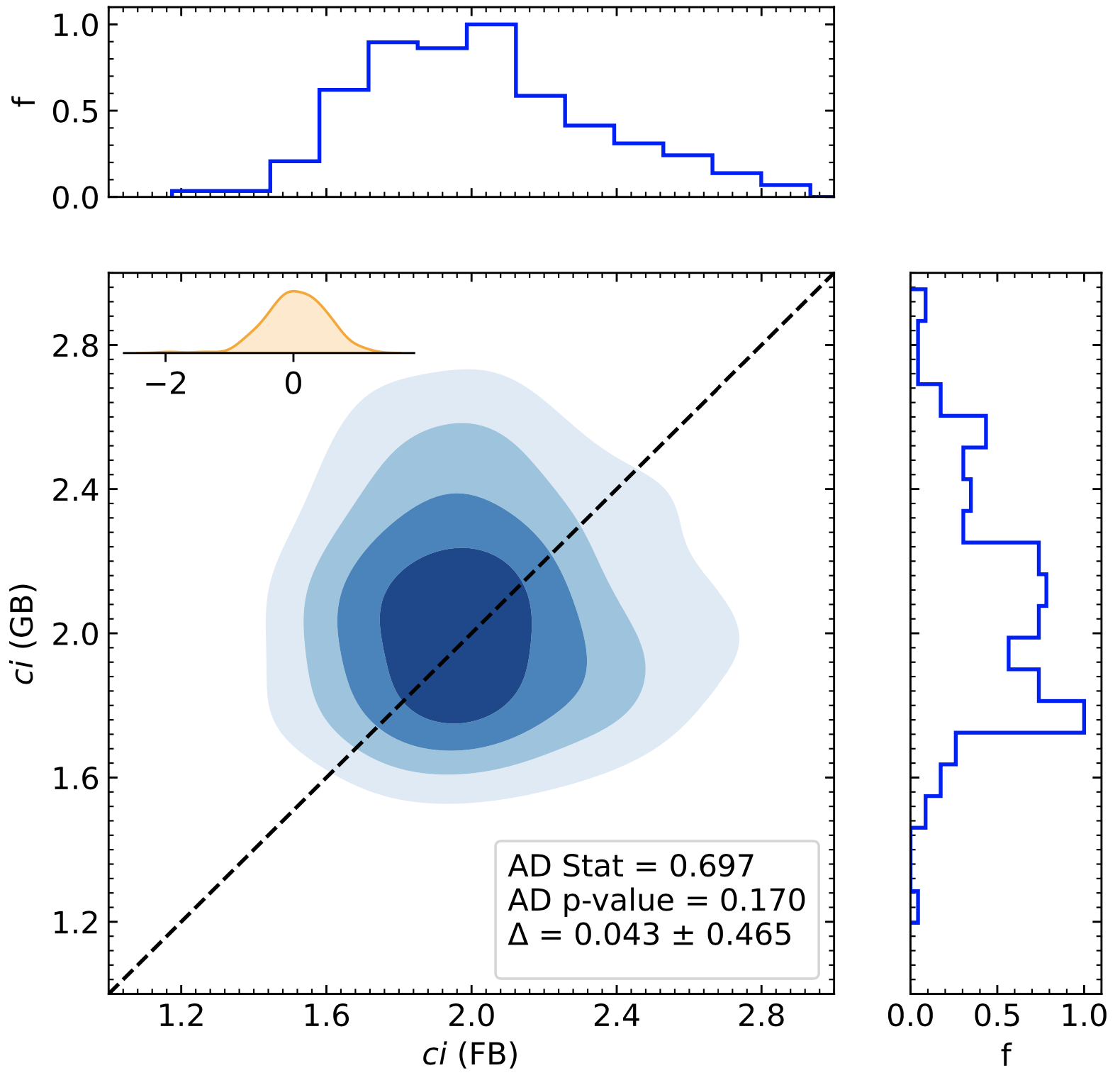}
	}
	\hspace{0.3cm}
	\subfigure{
		\includegraphics[width=0.46\linewidth]{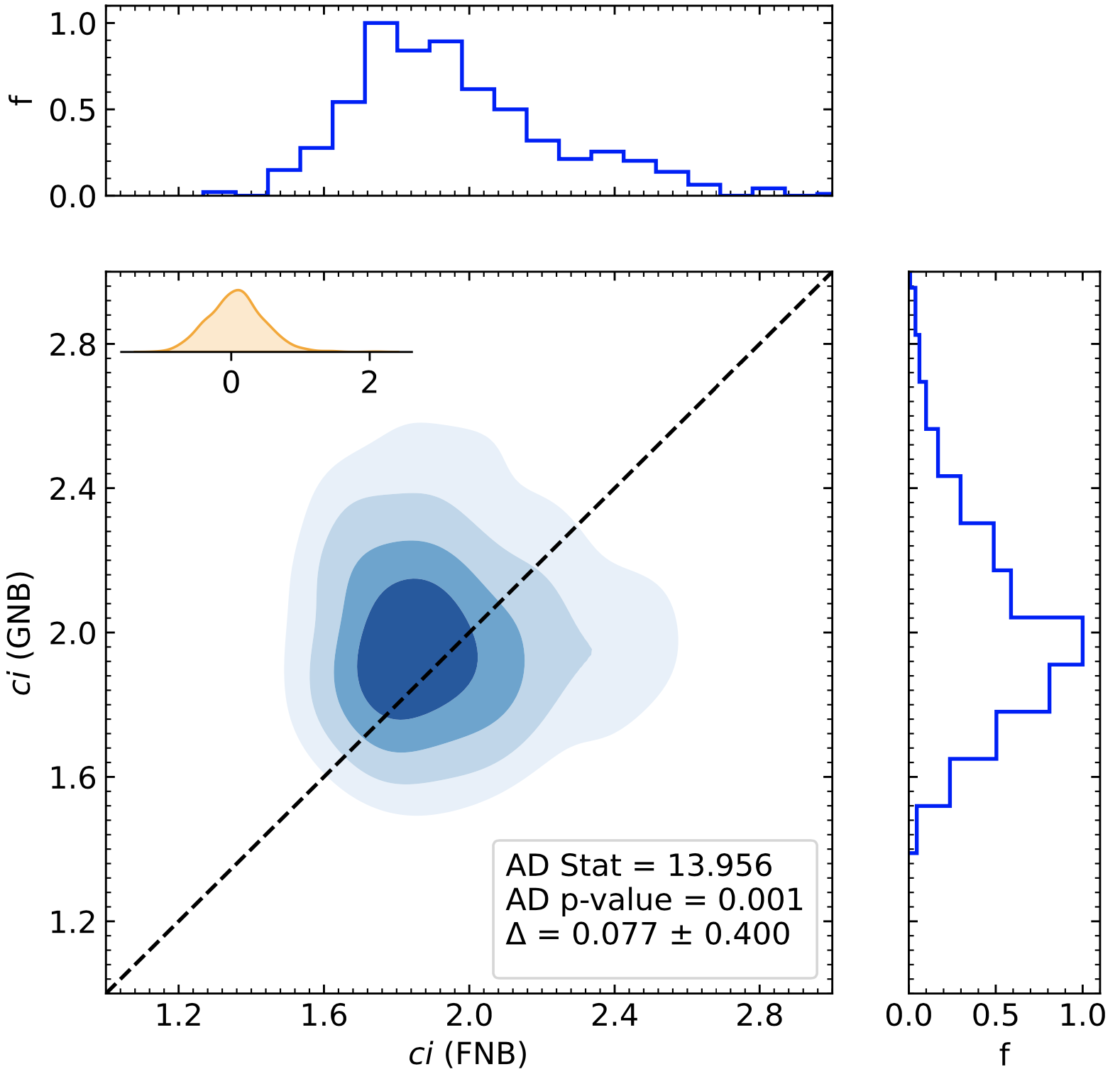}
	}
	\vspace{-0.2cm}
\caption{Comparison between concentration index ($ci$) distributions of 
barred (left panel) and unbarred (right panel) isolated (x-axis) and non-isolated
(y-axis) galaxies.}
\label{ci}
\end{figure}

\begin{figure}[h!]
	\subfigure{
		\includegraphics[width=0.46\linewidth]{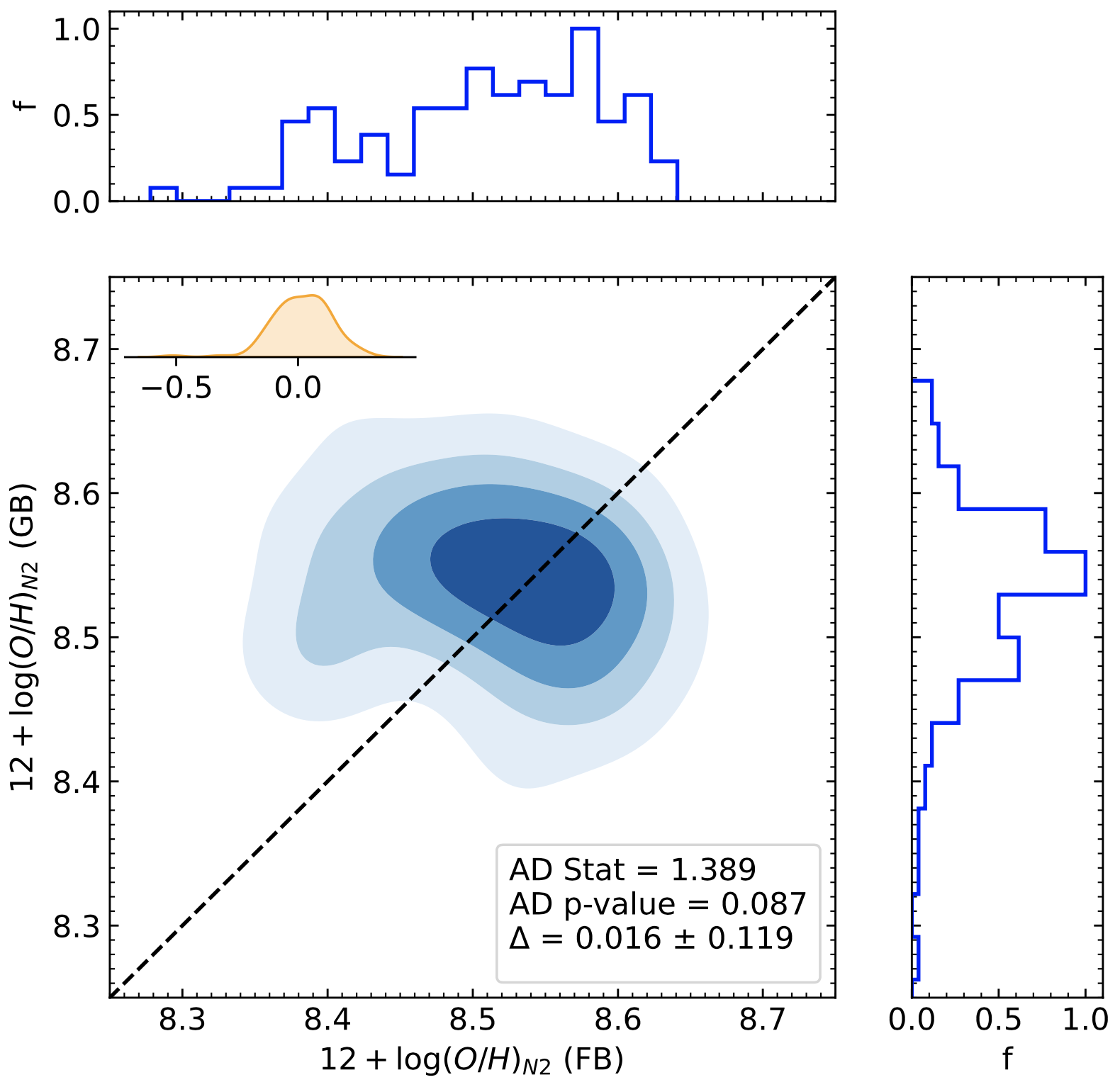}
	}
	\hspace{0.3cm}
	\subfigure{
		\includegraphics[width=0.46\linewidth]{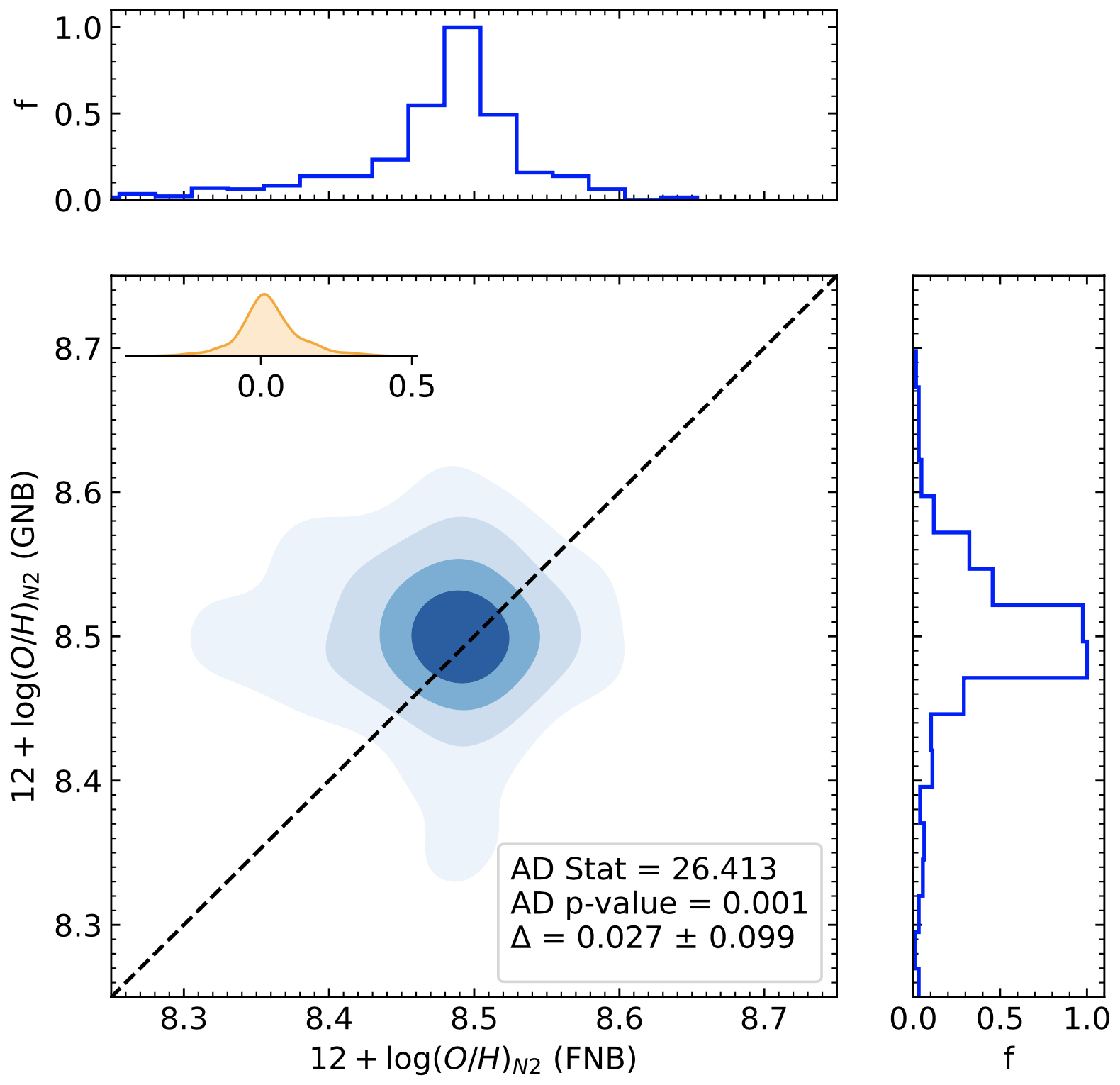}
	}
	\vspace{-0.2cm}
\caption{Comparison between gas phase metallicity using the calibrator $N2$ at 
the central region distributions of barred (left panel) and unbarred (right 
panel) isolated (x-axis) and non-isolated (y-axis) galaxies.}
\label{OHN2}
\end{figure}

In this section, we compare the selected properties between isolated and 
non-isolated 
galaxies obtained from the MaNGA survey for barred and unbarred galaxies as 
the tracer of the influence of bars on the environmental dependence of galaxy 
properties. Comparison of sets of physical properties of barred (left panel) 
and unbarred (right panel) galaxies are shown in Figs.~\ref{M},~\ref{SF},
\ref{GR},~\ref{ci}, and \ref{OHN2}. Each panel shows the density distribution 
as a set of contours filled in isolated and non-isolated parameter maps. Each 
successive 
line encloses the points of $80\%$, $60\%$, $40\%$ and $20\%$, respectively. 
A one-to-one relationship is indicated by the black dotted line.
The orange upper left inset plot located in the top-left corner of 
each figure represents the distribution of the differences in the properties
estimated between the isolated and non-isolated samples. For example, in the 
left panel of Fig.~\ref{M}, the x-axis of the inset shows the quantity 
\( \Delta = \log_{10} M_{*} (\text{GB}) - \log_{10} M_{*} (\text{FB}) \), 
which quantifies the offset in stellar masses on a galaxy-by-galaxy basis.
Here, GB stands for barred non-isolated galaxies and FB for barred isolated 
galaxies. This difference is expressed in dex, while the y-axis corresponds 
to the probability density and is unitless, indicating the relative frequency 
of galaxies exhibiting a given value of \( \Delta \). The shape of the 
distribution provides insight into the systematic differences between the 
two stellar mass estimates. In this case, the distribution is centred slightly 
above zero, consistent with the reported mean offset of 
\( \Delta = 0.242 \pm 0.745 \), suggesting that, on average, the GB stellar 
masses are marginally higher than those from FB.
The side panels show the 
normalised distributions for each sample where $f$ represent the normalized 
frequency. The Anderson-Darling p-value, the Anderson-Darling statistic, and 
the mean difference ($\Delta$) with its corresponding dispersion are observed 
in the shown legend in each panel of all figures for the selected properties. 
	
The galaxys' M$\star$ was estimated using photometry by means of the relation 
obtained from Ref.\ \cite{bell2000stellar}. Valid for the 
Ref.\ \cite{chabrier2003galactic} the initial mass function is given by 
equation,
\begin{equation}
  \log\left({M_{\ast,\text{phot}}}/{M_{\odot}}\right) = -\,0.95 + 1.58 (B - V) + 0.43 \times (4.82 - V_{\text{abs}}),
  \label{mass}
\end{equation}
where ${M_{\ast,\text{phot}}}/{M_{\odot}}$ is the photometry M$\star$ in solar 
mass unit, while the $(B - V)$ is colour index and $V_\text{abs}$ 
is the $V$ band absolute magnitude.
The star formation rate used in this study is derived from the dust-corrected 
H$\alpha$ luminosity ($L_{\text{H}\alpha}$) by employing equation, 
\begin{equation}
     \text{SFR} \, (M_{\odot} \, \text{yr}^{-1}) = 0.79 \times 10^{-41} \, L_{\text{H}\alpha} \, (\text{erg} \, \text{s}^{-1}).
     \label{star formation}
\end{equation}
The relationship was proposed by Ref.\ \cite{kennicutt1998star} for 
Ref.\ \cite{ salpeter1955luminosity}'s initial mass function. This SFR is the 
upper limit because in this derivation all the H$\alpha$ fluxes are summed up 
regardless of the detected ionization nature. This means that this calculation 
yields SFR even in non-ionising galaxies that can be directly related to 
recent star formation events \cite{cano2016spatially,sanchez2018sdss,
sanchez2021local}. 
The $g-r$ colour was extracted from the NSA catalogue obtained following 
the procedures outlined in Ref.\ \cite{sanchez2022sdss}. We used $R_{50}$ to 
represent the radius around $50\%$ and $R_{90}$ to represent the radius around 
$90\%$ of the Petrosian flux, respectively. We then calculated the structural 
parameter, the r-band concentration index as the ratio between the two radii 
given by $ci = R_{90} /R_{50}$. The two subsamples' distributions for $g-r$ 
colour and $ci$ in a non-isolated and isolated environment are shown in Figs.~\ref{GR} 
and \ref{ci}. 
	
For understanding the existence of different oxygen abundance calibrators as 
detailed in Refs.\ \cite{marino2013o3n2,vogt2015galaxy,perez2009impact,
sanchez2019characterizing,pilyugin2010new,kewley2008metallicity,
maiolino2008amaze,tremonti2004origin,dopita1996spectral,pena2012recalibration,
lopez2012eliminating,curti2020mass,ho2019machine,espinosa2022h}, 
in this study we adopted the estimate of the oxygen abundance from the 
N2-based calibrator proposed by Ref.\ \cite{marino2013o3n2} 
($12+\log(O/H)_{N2}$), which was obtained by using equation, 
\begin{equation}
	12 + \log \left(\frac{O}{H}\right) = 8.743 \, [\pm\, 0.027] + 0.462 \, [\pm\, 0.024] \times N2, \label{N2O}
\end{equation}
where
%\begin{equation}
	$N2 = \log \left([\text{N\,II}] \, \lambda 6583/\text{H}\alpha \right).$%\label{N2}
%\end{equation}
The results of this particular study have been shown in Fig.~\ref{OHN2}.
	
\subsection{Galaxy properties relationship}
\begin{figure}[t!]
	\subfigure{
		\includegraphics[width=0.47\linewidth]{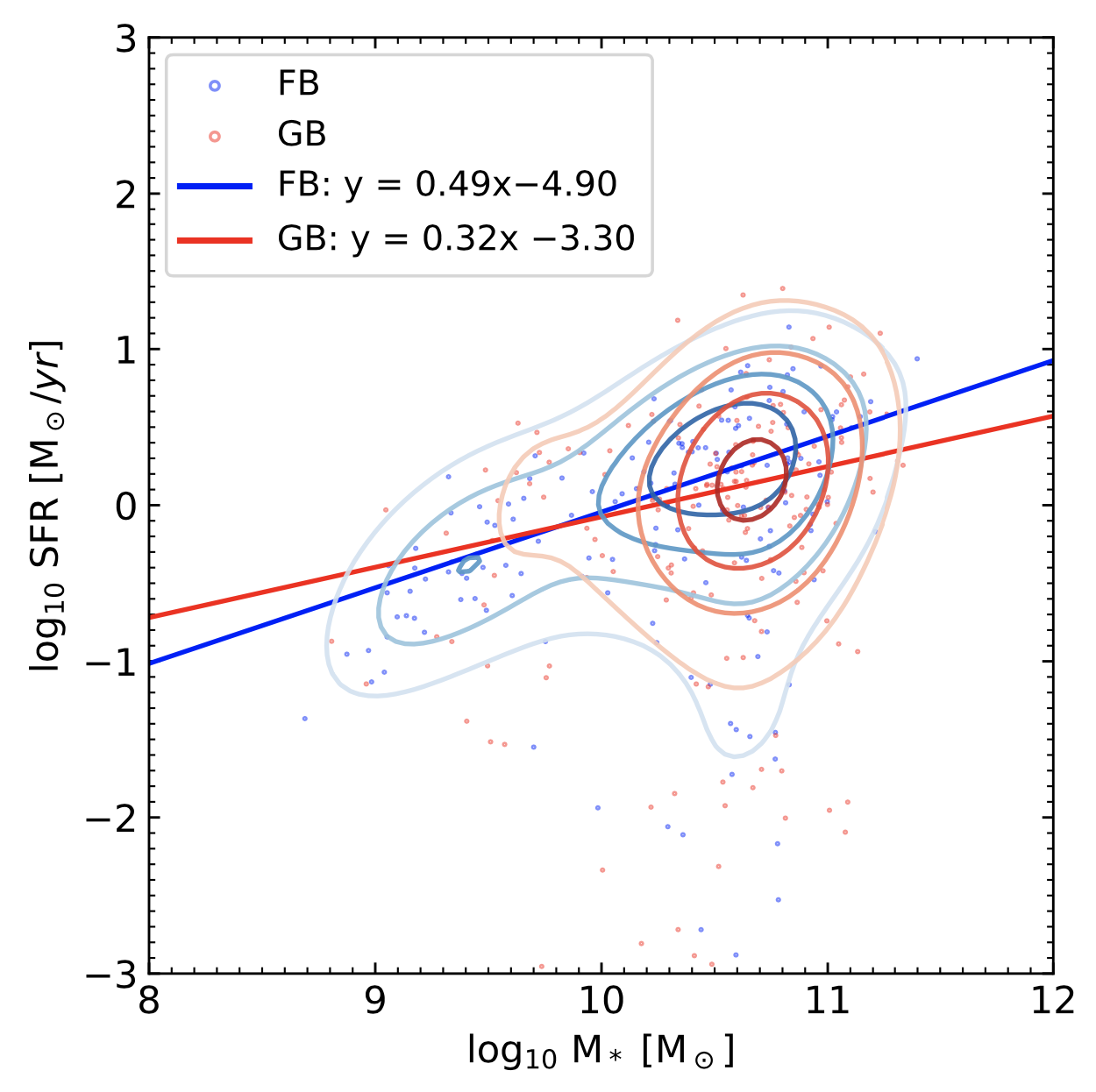}
	}
		\hspace{0.3cm}
	\subfigure{
		\includegraphics[width=0.47\linewidth]{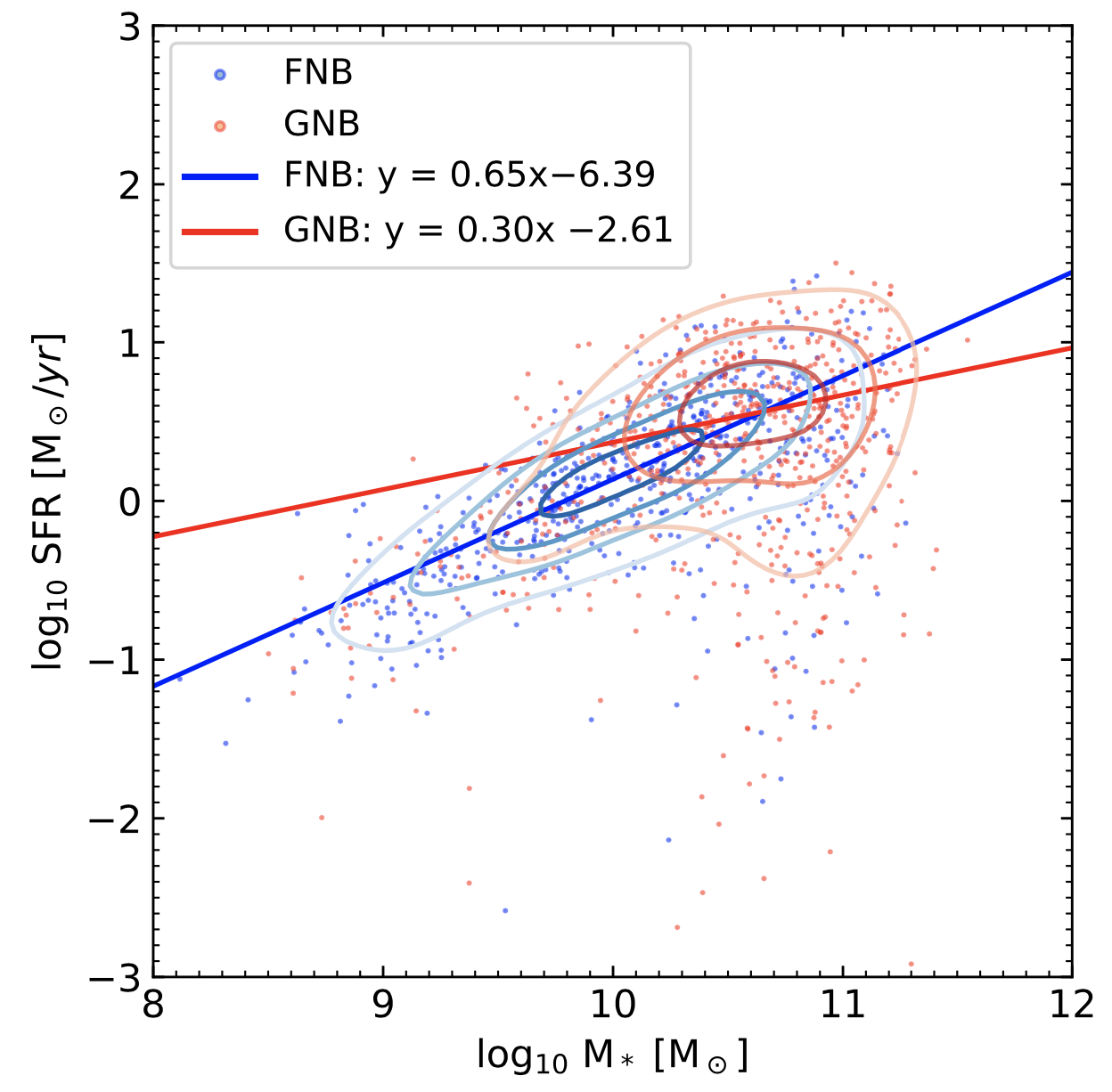}
	}
	\vspace{-0.2cm}
\caption{Variation of stellar mass against SFR for barred (left panel) and 
unbarred (right panel) galaxies. The contour levels in this plot and other 
similar plots represent the density levels from the probability density 
function (PDF) estimated using Kernel Density Estimation (KDE). 
Each successive line encloses the points of $80\%$, $60\%$, $40\%$ and $20\%$  in 
both isolated (F), non-isolated (G) in blue and red colours, respectively  for  barred (B)  
and unbarred (NB) galaxies.}
\label{MSF}
\end{figure}

\begin{figure}[h!]
	\subfigure{
		\includegraphics[width=0.47\linewidth]{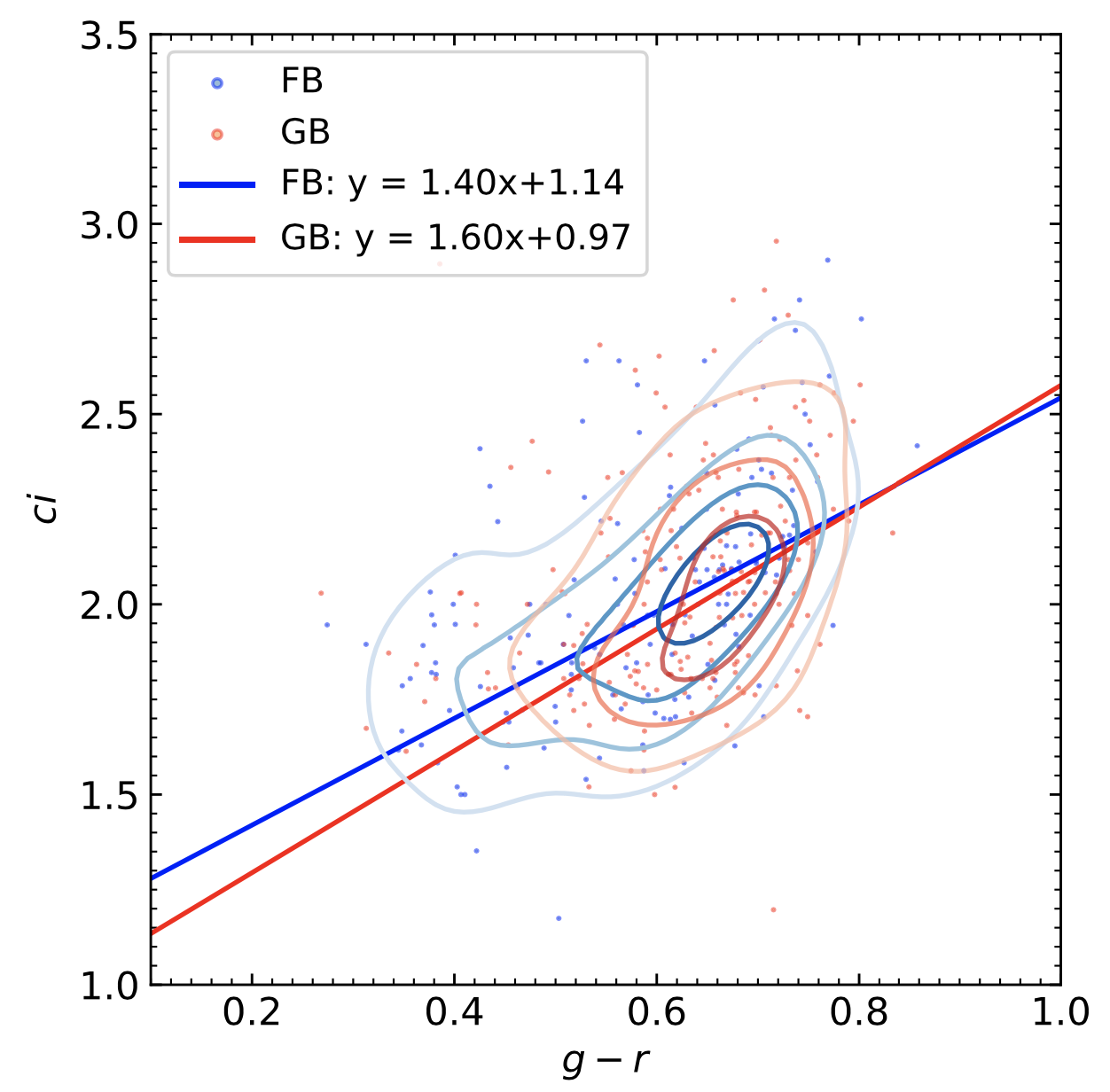}
	}
	\hspace{0.3cm}
	\subfigure{
		\includegraphics[width=0.47\linewidth]{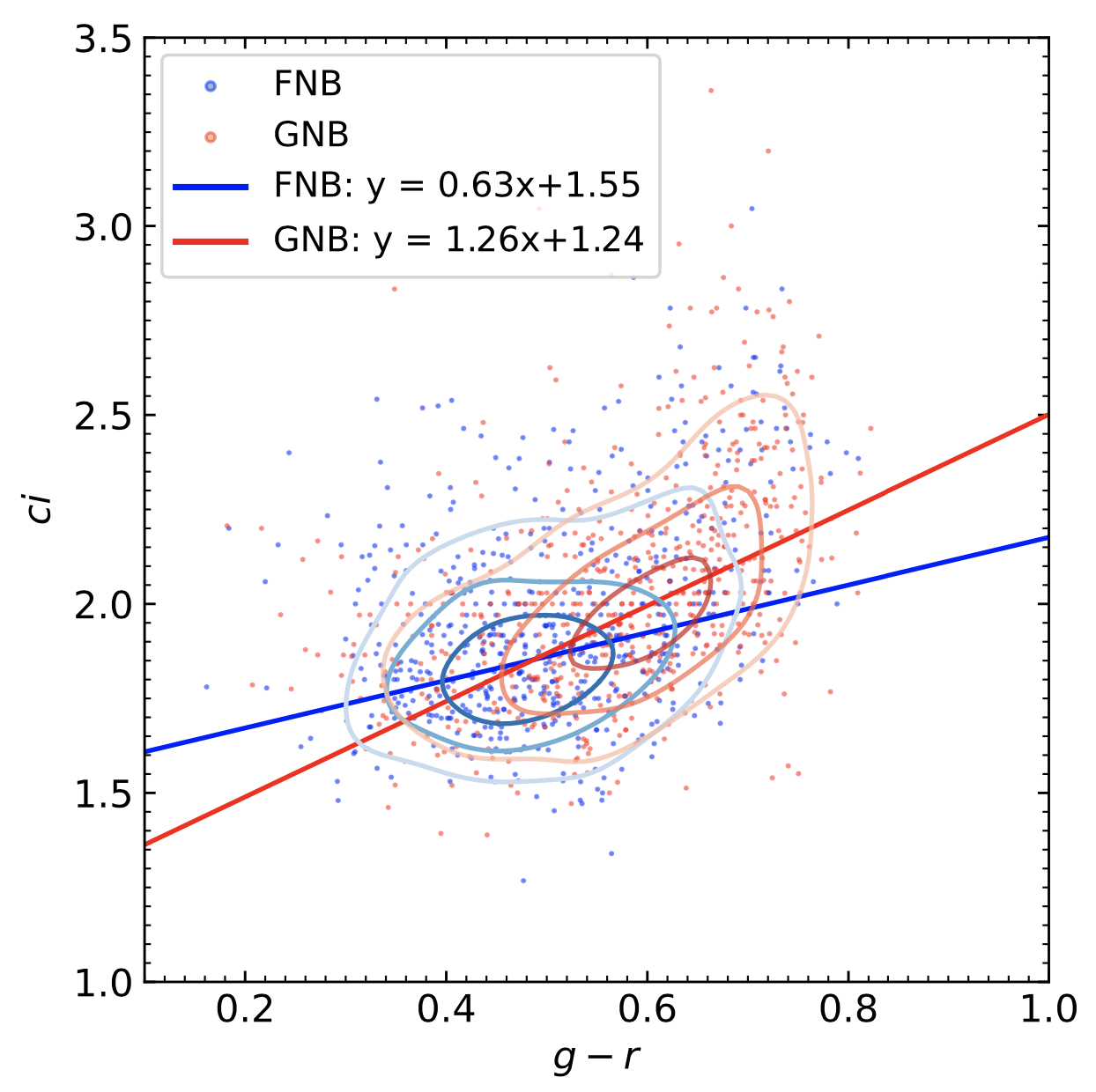}
	}
	\vspace{-0.2cm}
\caption{Variation of $g-r$ colour against concentration index ($ci$) for 
barred (left panel) and unbarred (right panel) galaxies. }
\label{GRC}
\end{figure}

\begin{figure}[h!]
	\subfigure{
		\includegraphics[width=0.47\linewidth]{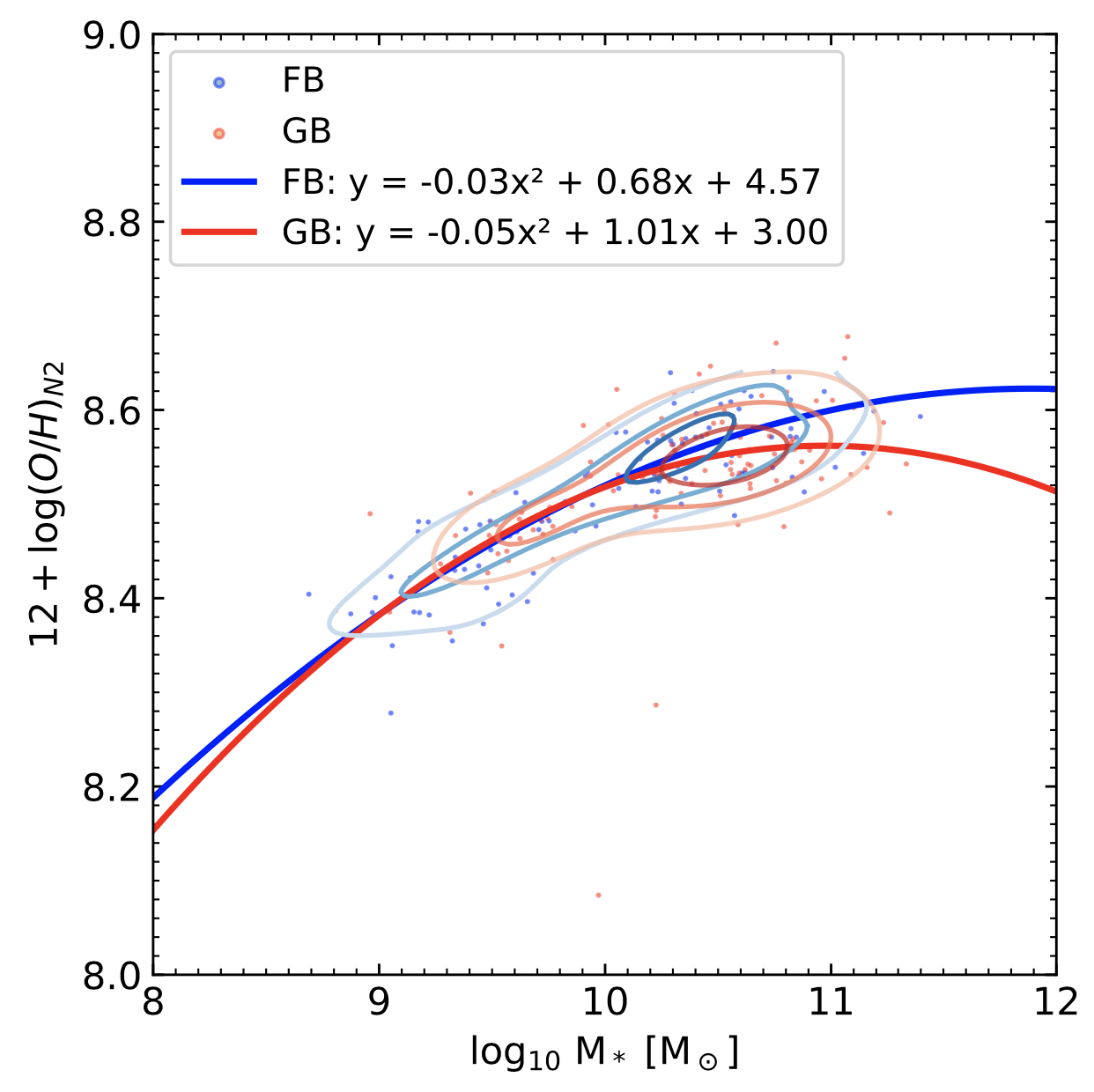}
	}
	\hspace{0.3cm}
	\subfigure{
		\includegraphics[width=0.47\linewidth]{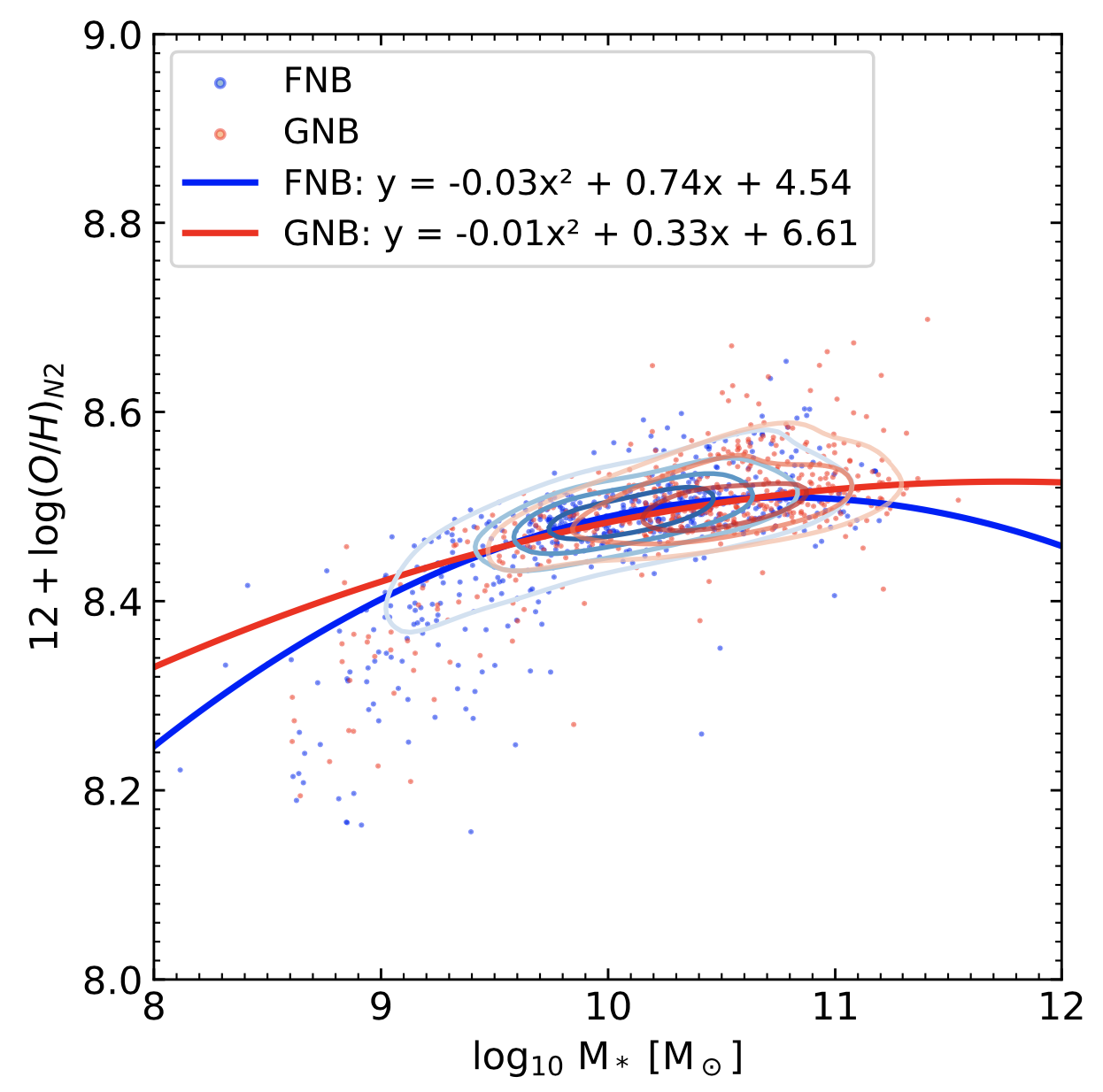}
	}
	\vspace{-0.2cm}
\caption{Variation of gas phase metallicity against stellar mass for barred 
(left panel) and unbarred (right panel) galaxies.}
\label{MM}
\end{figure}

In this section we derive the fundamental relationships between the properties 
of galaxies obtained in the previous section, aiming to observe if there is 
any significant difference in the environmental dependence of barred galaxies 
when compared to unbarred ones. For this purpose, we plot Figs.~\ref{MSF}, 
\ref{GRC}, and \ref{MM} that show the variation in stellar mass against SFR, 
$g-r$ colour against concentration index ($ci$) and Stellar mass against gas 
phase metallicity, respectively.
	
From the left panel of Fig.~\ref{MSF} for barred galaxies by performing the 
regression analysis the general equations for the best-fitted lines to the 
isolated and non-isolated galaxies' stellar masses against SFRs plots are respectively 
can be obtained as 
\begin{align}
	\log_{10}(\mbox{SFR}) & = 0.49\pm 0.11\,\log_{10}(M_\star) - 4.9\pm1.10,
	\label{MSF1}\\[5pt]
	\log_{10}(\mbox{SFR}) & = 0.32\pm 0.16\,\log_{10}( M_\star) - 3.30\pm 1.71.
\label{MSG1}
\end{align}
From the right panel of Fig.~\ref{MSF} for unbarred galaxies by performing the 
regression analysis the general equations for the best-fitted lines to the 
isolated and non-isolated galaxies' stellar masses against SFRs plots are respectively
can be obtained as 
\begin{align}
	\log_{10}(\mbox{SFR}) & = 0.65\pm 0.03\,\log_{10}(M_\star) - 6.39\pm0.33,
	\label{MSF2}\\[5pt]
	\log_{10}(\mbox{SFR}) & = 0.30\pm 0.05\,\log_{10}( M_\star) - 2.61\pm 0.51.
	\label{MSG2}
	\end{align}

From the left panel of Fig.~\ref{GRC} for barred galaxies by performing the 
regression analysis the general equations for the best-fitted lines to the 
isolated and non-isolated galaxies' $g-r$ colours against concentration indices ($ci$s) 
plots are respectively obtained as  
\begin{align}
	ci & = 1.40\pm 0.19\,(g-r) - 1.14\pm0.21,
	\label{GRCF1}\\[5pt]
	ci & = 1.60\pm 0.20\,(g-r) - 0.97\pm 0.21.
	\label{GRCG1}
\end{align}
From the right panel of Fig.~\ref{GRC} for unbarred galaxies by performing 
the regression analysis the general equations for the best-fitted lines to 
the isolated and non-isolated galaxies' $g-r$ colours against concentration indices 
($ci$s) are respectively obtained as 
\begin{align}
	ci & = 0.63\pm 0.09\,(g-r) - 1.55\pm0.05,
	\label{GRCF2}\\[5pt]
	ci & = 1.26\pm 0.09\,(g-r) - 1.24\pm 0.05.
	\label{GRCG2}
\end{align}
		
From the left panel of Fig.~\ref{MM} for barred galaxies by performing the 
regression analysis the general equations for the best-fitted lines to 
the isolated and non-isolated galaxies' gas phase metallicities against stellar masses 
are respectively found as 
\begin{align}
	12+\log_{10}(\mbox{O/H}_{N2}) & = -\,0.03\pm 0.01\,\log_{10}(M_\star)^{2} +0.68\pm 0.15\,\log_{10}(M_\star) + 4.57\pm0.73,
	\label{MMF1}\\[5pt]
	12+\log_{10}(\mbox{O/H}_{N2}) & = -\,0.05\pm 0.02\,\log_{10}(M_\star)^{2} +1.01\pm 0.33\,\log_{10}(M_\star) + 3.00\pm1.67
	\label{MMG1}
\end{align}
From the right panel of Fig.~\ref{MM} for unbarred galaxies by performing the 
regression analysis the general equations for the best-fitted line to the 
isolated and non-isolated galaxies' gas phase metallicities against stellar masses 
are respectively found as 
\begin{align}
	12+\log_{10}(\mbox{O/H}_{N2}) & = -\,0.03\pm 0.01\,\log_{10}(M_\star)^{2} +0.74\pm 0.10\,\log_{10}(M_\star)+ 4.54\pm0.49,
	\label{MMF2}\\[5pt]
	12+\log_{10}(\mbox{O/H}_{N2}) & = -\,0.01\pm 0.01\,\log_{10}(M_\star)^{2} +0.33\pm 0.09\,\log_{10}(M_\star) + 6.61\pm0.48
	\label{MMG2}
\end{align}
		
\section{Discussion}
	\label{secIV}
The stellar masses shown in the left panel of Fig.~\ref{M} for barred galaxies 
present a systematic offset ($\Delta$) of $\sim 0.2$ dex between isolated and 
non-isolated galaxies' datasets where M$\star$ of the non-isolated is slightly larger than 
the isolated. The observed difference is much less when compared to the spread of 
the M$\star$ characterised by the standard deviation given by 
$\sigma_{M\star} \sim 0.7$ indicating a one-to-one correspondence between isolated 
and non-isolated galaxies' M$\star$. The distributions of the two data sets indicate 
that they are the same, having the AD statistics of $\sim 0.4$ and p-value 
of $\sim 0.212$. A systematic offset of $\sim 0.4$ dex between the unbarred 
isolated and non-isolated galaxies' M$\star$ is shown in the right panel of 
Fig.~\ref{M}, where the non-isolated galaxies' masses are greater than the isolated 
ones. Comparing this difference with the spread between the two data sets 
characterised by the standard deviation of $\sigma_{M\star} \sim 0.8$, this 
difference is far smaller. The distributions of the two data sets indicate 
that they are different by having different AD statistics of $\sim 74.1$ 
and p-value of $\sim 0.001$. The presence of bars results in a decrease 
of the difference in masses between isolated and non-isolated galaxies by $\sim 50\%$. 
The large value of AD statistics and small p-value for unbarred galaxies 
indicate the strong dependence of M$\star$ on the environment when compared 
to barred galaxies. These facts imply that the presence of bars decreases the 
dependence of M$\star$ on the environment for the disc galaxies.
	
The left panel of Fig.~\ref{SF} shows a comparison of SFRs for barred galaxies 
obtained using the dust-corrected H$\alpha$ luminosity as already stated, 
where the spread is around a one-to-one relationship presenting a systematic 
offset of only $\sim 0.1$ dex between the barred isolated and non-isolated galaxies' 
SFRs, with the isolated galaxies' SFRs being slightly larger than the non-isolated ones. 
This difference is much smaller when compared to the spread between isolated and 
non-isolated galaxies characterised by the standard deviation of SFRs 
$\sigma_{S\!F\!R} \sim 1.5$. The distributions of the two data sets indicate 
that they are similar, having the AD statistics of $\sim 0.4$ and p-value of 
$\sim 0.2$. A consistent offset of $\sim 0.2$ dex between the two datasets is 
shown in the right panel of Fig.~\ref{SF} for the unbarred galaxies, where the 
non-isolated galaxies' SFRs are larger than the ones in isolated galaxies. Compared to 
the spread between these two sets of values, which is defined by the SFR 
standard deviation $\sigma_{S\!F\!R} \sim 0.9$, this difference is far smaller, 
again indicating a one-to-one relationship. However, the distributions of the 
two data 
sets indicate that they are different, having the AD statistics of $\sim 34.5$ 
and p-value of $\sim 0.001$. These results imply a strong dependence of 
SFRs for unbarred galaxies on the environment when compared to that of barred 
galaxies, indicating that the presence of bars decreases the dependence of SFRs on the environment.
	
Figs.~\ref{GR} and \ref{ci} show a one-to-one correspondence between isolated 
and non-isolated disc galaxies for both barred (left panels) and unbarred 
(right panels) ones. The barred samples show the offset of $\sim 0.04$ dex for 
$g-r$ colour and $ci$ while the unbarred samples show $\sim 0.1$ dex offset. 
These differences are smaller than the spread between the two sets of values 
characterised by standard deviations of $\sigma_{g-r} \sim 0.2$ and 
$\sigma_{ci} \sim 0.4$. We also performed the Anderson–Darling statistical 
test as in earlier cases for the $g-r$ and $ci$ distributions of two 
subsamples at isolated and non-isolated environments and found that the $g-r$ and $ci$ of 
unbarred galaxies strongly depend on environment with AD statistics of 
$\sim 52.4$ and $\sim 14$ respectively when compared to barred galaxies with 
AD statistics of $\sim 0.3$ and $\sim 0.7$ respectively. Similarly, the lower 
p-values ($0.001$ for both $g-r$ colour and $ci$) of unbarred samples when 
compared to barred ones ($0.3$ and $0.2$ for $g-r$ and $ci$, respectively) 
reinforce the observation that bars result in the decrease of the dependence 
of both $g-r$ colour and $ci$ on the environment. Almost all of the galaxies 
in this study are of the late-type, with $ci < 2.86$, which is realistic since 
the study involves spiral galaxies that are indeed late-type.  
	
The left panel of Fig.~\ref{OHN2} displays the gas phase metallicity of 
barred galaxies, which shows a systematic offset of $\sim 0.02$ dex between 
isolated and non-isolated barred galaxies' metallicity where the non-isolated's metallicity 
is slightly higher than the isolated ones. This difference is significantly 
smaller when compared to the spread between the two datasets as characterised 
by a standard deviation of $\sim 0.1$, indicating a one-to-one correspondence 
in metallicity between barred isolated and non-isolated galaxies. The distributions 
of the two data sets indicate that they are the same, having the AD statistics 
of $\sim 1.4$ and p-value of $\sim 0.09$. A systematic offset of $\sim 0.03$ 
dex for unbarred galaxies where the non-isolated metallicity is greater than the 
isolated ones is shown in the right panel of Fig.~\ref{OHN2}. In comparison with 
the spread between the two datasets, with a standard deviation of $\sim 0.1$, 
this difference is much smaller which implies that there is a one-to-one 
correspondence. The distributions of the two data sets indicate that they are 
different, having the AD statistics of $\sim 26.4$ and p-value of $\sim 0.001$. 
The large value of AD statistics and small p-value for metallicity indicate 
the strong dependence of unbarred galaxies on the environment when compared 
to barred ones. These facts imply that the presence of bars decreases the 
dependence of gas phase metallicity on the environment for disc galaxies. 
	
The difference in slopes between Eqs.~\eqref{MSF1} and \eqref{MSG1} is $0.17$ 
dex and the difference in intercepts is $1.60$, which are within the maximum 
errors in the measurements ($0.16$ and $1.71$ for the slope and intercept, 
respectively). These differences produce p-values of $0.403$ and $0.431$ in 
slope and intercept, respectively by t-test. Since these p-values are much 
greater than the standard statistics' p-value ($0.05$) these differences 
in slope and intercept are less significant which implies that the variation 
of M$\star$ and SFR for barred galaxies are not significantly influenced by 
the environment. On the other hand the difference in slopes between 
Eqs.~\eqref{MSF2} and \eqref{MSG2} is $0.35$ dex and the difference in 
intercept is $3.78$, which are much greater than the maximum errors in the 
measurement ($0.05$ and $0.51$ for the slope and intercept, respectively). 
These differences produce p-values of \num{2.027e-09} and \num{6.424e-10} in 
slope and intercept, respectively. Since these p-values are much less than 
the standard statistics' p-value ($0.05$) the differences in slope and 
intercept are much more significant which implies that the variation of 
M$\star$ and SFR for unbarred galaxies is affected by the environment.
	
The difference in slopes between Eqs.~\eqref{GRCF1} and \eqref{GRCG1} is 
$0.20$ dex and the difference in intercepts is $0.17$. These differences are 
within the maximum errors associated with the slope ($\sim 0.20$) and 
intercept ($\sim 0.21$). Furthermore, the p-values produced by these 
differences are $\sim 0.478$ and $\sim 0.338$ for the slope and intercept, 
respectively by t-test. Since these p-values are much greater than the 
standard statistics' p-value ($0.05$) these differences in slope and intercept 
are less significant which implies that the variation of $g-r$ colours and 
concentration indices ($ci$s) for barred galaxies are not significantly 
influenced by the environment. On the other hand the difference in slopes 
between Eqs.~\eqref{GRCF2} and \eqref{GRCG2} is $0.63$ dex and the difference 
in intercepts is $0.31$, which are much larger than the maximum errors in 
the measurement ($\sim 0.09$ and $\sim 0.05$ for the slope and intercept, 
respectively). Moreover, these differences produce p-values of 
\num{5.176e-07} and \num{9.255e-06} in slope and intercept, respectively. 
Since these p-values are much less than the standard statistics' p-value 
($0.05$) the differences in slope and intercept are much more significant 
which implies that the variation of $g-r$ colour and concentration index 
($ci$) for unbarred galaxies is affected by the environment.
	
The differences in quadratic coefficients (curvature), linear coefficients 
(slopes), and intercepts between Eqs.~\eqref{MMF1} and \eqref{MMG1} are 
$\sim 0.02$ dex, $\sim 0.33$ dex, and $\sim 1.57$ dex, respectively. 
These differences are within the maximum errors associated with the curvature, 
slope and intercept given by $\sim 0.02$, $\sim 0.33$  and $\sim 1.67$, 
respectively. Furthermore, the p-values produced by these differences are 
$\sim 0.323$, $\sim 0.358$ and $\sim 0.392$ for the curvature, slope and 
intercept, respectively using the t-test. Since these p-values are much greater 
than the standard statistics' p-value ($0.05$) the differences in curvature, 
slope and intercept are less significant which implies that the variation of 
stellar mass against gas phase metallicity for barred galaxies are not 
significantly influenced by the environment. On the other hand, the difference 
in curvatures, slopes, and  intercepts  between Eqs.~\eqref{GRCF2} and 
\eqref{GRCG2} are $\sim 0.02$ dex, $\sim 0.41$ dex and $\sim 2.07$ dex, 
respectively, which are much larger than the maximum errors in the measurement 
($\sim 0.01$, $\sim 0.10$ and $\sim 0.49$). Furthermore, these differences 
produce p-values of \num{2.230e-03}, \num{2.368e-03}, \num{2.580e-03} in 
curvature, slope and intercept, respectively. Since the p-values are less 
than the standard p-value in statistics ($0.05$), the differences in 
curvature, slope and intercept are significant which implies that the 
variation of stellar mass against gas phase metallicity for unbarred galaxies 
is affected by the environment.
	
The results are consistent with the findings of Ref.\ 
\cite{goto2003morphology}, who observed that a large fraction of spirals 
in dense environment have early-type morphology than the isolated similar to 
the evidence from Fig.\ref{GRC} whereby non-isolated galaxies have high 
concentration indices (early type) when compared to isolated for both barred 
and unbarred galaxies. The study agrees with the results that mostly early-type 
spirals are found in group environments compared to late-type 
\cite{goto2003morphology,aguerri2004environmental}, similarly we 
observed most barred galaxies to exist in groups 
($\sim 56\%$) when compared to unbarred ($\sim 51\%$), in this case barred, 
unbarred galaxies resemble early-type, late-type, respectively 
based on the findings that bars forms over time \cite{lokas2019buckling,ansar2024bar,kwak2017origin,
lopez2024unveiling,izquierdo2022disc}. From Figs.\ref{GRC} non-isolated galaxies 
have higher $g-r$ colour (most redder) when compare to isolated supporting 
the results from Refs.\ \cite{wolf2009stages,bamford2009galaxy,masters2010galaxy} 
that the blue spiral resides outside the group environments than red spiral galaxies. 
The M$\star$ against SFR and $g-r$ against $ci$ relationships as shown in 
Figs.\ref{MSF}, \ref{GRC} are well fitted by a linear line, the most widely 
proposed relation in a number of works for the given mass ranges 
E.g.~in Refs.\ \cite{elbaz2007reversal,speagle2014highly,leslie2015quenching,
	daddi2007multiwavelength,yuan2010role,
rich2011galaxy,leslie2015quenching,schawinski2007observational,
whitaker2012star} while the M$\star$ against gas phase metallicity 
relationships as shown in Fig.\ref{MM} is 
well fitted by a polynomial supporting the study by 
Ref.\ \cite{tremonti2004origin}.

Our analysis suggests that bars moderate the influence of the environment on 
key galaxy properties, including stellar mass, SFR, $g-r$ colour, 
concentration index, and gas-phase metallicity. When comparing barred and 
unbarred disc galaxies in isolated and non-isolated environments, we find that 
barred galaxies tend to exhibit weaker environmental trends across these 
parameters. In particular, barred galaxies show less sensitivity to the 
environment for their star formation rates, colours, concentration index 
and gas-phase metallicity when compared to unbarred, suggesting that internal 
secular processes may dominate over external mechanisms in shaping these 
structural and chemical properties. These results imply that bars may act as 
internal regulators, partially decoupling galaxy evolution from environmental 
dependence, particularly in the transition from isolated to non-isolated 
environments.
	
\section{Summary and Conclusion}
\label{secV}
Using a volume-limited sample obtained from the MaNGA survey we investigated 
the influence of bars on the environmental dependence of disc galaxies' 
physical properties. We selected the galaxies with bars using the criteria 
\eqref{c1}, \eqref{c2}, \eqref{c3}, and \eqref{c4} then unbarred galaxies, 
using criteria \eqref{c1}, \eqref{c2}, \eqref{c3}, and \eqref{c5} to obtain 
a total of  $356$ and $1180$ for the galaxies with bars and without bars 
with samples shown in Figs.~\ref{B} and \ref{NB}, respectively.
To quantify the galaxy's environmental effect we used the 
Galactic Environment for the MaNGA Value Added Catalogue (GEMA-VAC) 
obtained by using the methods described in Ref.\ \cite{argudo2015catalogues,
etherington2015measuring,wang2016elucid} which is a volume-limited sample up 
to $z < 0.15$ whereby the galaxies are assigned in groups using the 
halo-based group finder used by Ref.\ \cite{yang2007galaxy}. The galaxies 
which are alone in the group (Group size (GS) $=1$) are referred to as isolated 
and the galaxies with at least one neighbour (Group size (GS) $\geq2$) are 
named non-isolated  galaxies. A total number of  $158$ $(44.38\%)$ and $198$ 
$(55.62\%)$ isolated and non-isolated barred galaxies were obtained. Similarly, a total 
number of $572$ $(48.47\%)$ and $608$ $(51.53\%)$ isolated and non-isolated unbarred 
galaxies were obtained. These samples were used to compare the M$\star$, SFR, 
colour, r-band concentration index ($ci$) and gas phase metallicity between 
isolated and 
non-isolated environments as shown by Figs.~\ref{M}, \ref{SF}, \ref{GR}, \ref{ci} and 
\ref{OHN2}, respectively. Then these were used to investigate if there is the 
existence of any difference between barred and unbarred galaxies. The  
M$\star$ against SFR, $g-r$ against $ci$, and M$\star$ against gas phase 
metallicity were studied in Figs.~\ref{MSF}, \ref{GRC} and \ref{MM}, 
respectively. Together with the already established results, this study 
revealed the following:
\begin{itemize}
\item A one-to-one correspondence of M$\star$, SFR, $g-r$ colour, $ci$ and gas 
phase metallicity between isolated and non-isolated galaxies was observed for 
both barred and unbarred galaxies.
\item For unbarred galaxies, the M$\star$, SFR, $g-r$ colour, $ci$ and gas 
phase metallicity exhibit a strong correlation with the environment, while 
these same properties for barred galaxies display a notably weaker 
environmental dependence.
\item The slope and intercept of M$\star$ against SFR relation of barred 
galaxies are weakly dependent on the environment while for unbarred there is a 
strong dependence.
\item There is a significant difference in the slopes and intercepts 
of $g-r$ colour against $ci$ relation between isolated and non-isolated for unbarred 
galaxies, while for barred the differences are insignificant.
\item The insignificant difference in curvatures, slopes and intercepts of 
M$\star$ against gas phase metallicity when isolated and non-isolated barred galaxies 
are compared were observed while for unbarred galaxies the differences are 
significant.
\end{itemize}
The study revealed that the presence of bars in disc galaxies decreases the 
dependence of analysed properties and their relations on the environment. The 
study emphasises on bar consideration in the investigation of the environmental 
dependence of disc galaxies' properties.
	
\section*{Acknowledgements} PP acknowledges support from The Government of Tanzania 
through the India Embassy, Mbeya University of Science and Technology (MUST) 
for Funding and SDSS for providing data. UDG is thankful to the Inter-University 
Centre for Astronomy and Astrophysics (IUCAA), Pune, India for 
the Visiting 
Associateship of the institute. Funding for the Sloan Digital Sky Survey IV has been provided by 
the Alfred P. Sloan Foundation, the U.S. Department of Energy Office of Science, and the 
Participating Institutions. SDSS-IV acknowledges support and resources from the 
Center for High-Performance Computing at the University of Utah. SDSS-IV is managed by 
the Astrophysical Research Consortium for the Participating Institutions of the SDSS Collaboration, 
including the Brazilian Participation Group, the Carnegie Institution for Science, 
Carnegie Mellon University, Center for Astrophysics | Harvard \& Smithsonian, the 
Chilean Participation Group, the French Participation Group, Instituto de Astrofísica de Canarias, 
The Johns Hopkins University, Kavli Institute for the Physics and 
Mathematics of the Universe (IPMU)/University of Tokyo, 
the Korean Participation Group, Lawrence Berkeley National Laboratory, 
Leibniz Institut für Astrophysik Potsdam (AIP), Max-Planck-Institut für Astronomie (MPIA Heidelberg), 
Max-Planck-Institut für Astrophysik (MPA Garching), Max- Planck-Institut für Extraterrestrische Physik (MPE), 
National Astronomical Observatories of China, New Mexico State University, New York University, 
University of Notre Dame, Observatário Nacional/MCTI, The Ohio State University, 
Pennsylvania State University, Shanghai Astronomical Observatory, United Kingdom Participation Group, 
Universidad Nacional Autónoma de México, University of Arizona, University of Colorado Boulder, 
University of Oxford, University of Portsmouth, University of Utah, University of Virginia, 
University of Washington, University of Wisconsin, Vanderbilt University, and Yale University.

\end{document}